\documentclass[
	aps,
	pra,
	twocolumn,
	reprint,
	superscriptaddress,
	floatfix,
	citeautoscript,
	longbibliography,
]{revtex4-2}

\usepackage[T1]{fontenc}
\usepackage[utf8]{inputenc}

\usepackage{newtxtext}
\usepackage{newtxmath}

\usepackage{chemformula}
\usepackage{siunitx}
\DeclareSIUnit\amu{u}
\DeclareSIUnit\atom{atom}
\DeclareSIUnit\stoppingunit{SU}
\DeclareSIUnit\torr{Torr}

\usepackage{graphicx}
\graphicspath{{./figures}}

\usepackage[
	unicode,
	colorlinks = true,
	allcolors = blue,
]{hyperref}

\usepackage{cleveref}
\usepackage{microtype}

\usepackage{glossaries}
\glsdisablehyper

\newacronym{1d}{1D}{one-dimensional}
\newacronym{2d}{2D}{two-dimensional}
\newacronym{3d}{3D}{three-dimensional}

\newacronym{ac}{AC}{alternating current}
\newacronym{afm}{AFM}{atomic force microscopy}
\newacronym{alc}{ALC}{avoided level crossing}
\newacronym{api}{API}{application programming interface}
\newacronym{ariel}{ARIEL}{Advanced Rare Isotope Laboratory}
\newacronym{arpes}{ARPES}{angle-resolved photoemission spectroscopy}
\newacronym{atp}{ATP}{adenosine triphosphate}

\newacronym[sort={b-NMR}]{bnmr}{\ensuremath{\beta}-NMR}{\ensuremath{\beta}-detected nuclear magnetic resonance}
\newacronym[sort={b-NMR2}]{bnmr2}{\ensuremath{\beta}-NMR}{\ensuremath{\beta}-radiation-detected NMR}
\newacronym[sort={b-NQR}]{bnqr}{\ensuremath{\beta}-NQR}{\ensuremath{\beta}-detected nuclear quadrupole resonance}
\newacronym{bca}{BCA}{binary collision approximation}
\newacronym{bcc}{BCC}{body-centred cubic}
\newacronym{bcp}{BCP}{buffered chemical polishing}
\newacronym{bcs}{BCS}{Bardeen-Cooper-Schrieffer}
\newacronym{bpp}{BPP}{Bloembergen-Purcell-Pound}
\newacronym{bsc}{BSC}{\ch{Bi2Se3:Ca}}
\newacronym{btm}{BTM}{\ch{Bi2Te3:Mn}}
\newacronym{bts}{BTS}{\ch{Bi2Te2Se}}

\newacronym{camp}{CAMP}{control and monitor program}
\newacronym{ccd}{CCD}{charge-coupled device}
\newacronym{cdw}{CDW}{charge density wave}
\newacronym{cgs}{CGS}{centimetre-gram-second system of units}
\newacronym{cmms}{CMMS}{Centre for Molecular and Materials Science}
\newacronym{codata}{CODATA}{Committee on Data for Science and Technology}
\newacronym{cpu}{CPU}{central processing unit}
\newacronym{create}{CREATE}{Collaborative Research and Training Experience Program}
\newacronym{cw}{CW}{continuous wave}

\newacronym{daq}{DAQ}{data acquisition}
\newacronym{dc}{DC}{direct current}
\newacronym{dft}{DFT}{density functional theory}
\newacronym{dos}{DOS}{density of states}
\newacronym{dqt}{DQT}{double-quantum transition}

\newacronym{efg}{EFG}{electric field gradient}
\newacronym{emim-ac}{EMIM-Ac}{1-ethyl-3-methylimidazolium acetate}
\newacronym{emim-dca}{EMIM-DCA}{1-ethyl-3-methylimidazolium dicyanamide}
\newacronym{ep}{EP}{electro-polishing}
\newacronym{epr}{EPR}{electron paramagnetic resonance}
\newacronym{erda}{ERDA}{elastic recoil detection analysis}
\newacronym{endor}{ENDOR}{electron nuclear double resonance}
\newacronym{epics}{EPICS}{Experimental Physics and Industrial Control System}
\newacronym{esr}{EPR}{electron spin resonance}

\newacronym{fcc}{FCC}{face-centred cubic}
\newacronym{fft}{FFT}{fast Fourier transform}
\newacronym{fom}{FoM}{figure of merit}
\newacronym{fwhm}{FWHM}{full width at half maximum}

\newacronym{gga}{GGA}{generalized gradient approximation}

\newacronym{hb}{HB}{hole-burning}
\newacronym{hfqs}{HFQS}{high-field \ensuremath{Q} slope}
\newacronym{hv}{HV}{high-voltage}
\newacronym{hwhm}{HWHM}{half width at half maximum}

\newacronym{iaea}{IAEA}{International Atomic Energy Agency}
\newacronym{icru}{ICRU}{International Commission on Radiation Units and Measurements}
\newacronym{il}{IL}{ionic liquid}
\newacronym{is}{IS}{impedance spectroscopy}
\newacronym{isac}{ISAC}{isotope separator and accelerator}
\newacronym{isol}{ISOL}{isotope separation online}
\newacronym{isosim}{IsoSiM}{Isotopes for Science and Medicine}

\newacronym{lcao}{LCAO}{linear combination of atomic orbitals}
\newacronym{lda}{LDA}{local density approximation}
\newacronym{led}{LED}{light-emitting diode}
\newacronym{leis}{LEIS}{low-energy ion scattering}
\newacronym{lib}{LIB}{lithium-ion battery}
\newacronym{lsat}{LSAT}{\ch{(La,Sr)(Al,Ta)O3}}

\newacronym{mas}{MAS}{magic angle spinning}
\newacronym{mpms}{MPMS}{magnetic property measurement system}
\newacronym{mbe}{MBE}{molecular beam epitaxy}
\newacronym{md}{MD}{molecular dynamics}
\newacronym{midas}{MIDAS}{Maximum Integrated Data Acquisition System}
\newacronym{mit}{MIT}{metal-insulator transition}
\newacronym{mnr}{MNR}{Meyer-Neldel rule}
\newacronym{mqt}{mqt}{multi-quantum transition}
\newacronym{mud}{MUD}{muon data}
\newacronym{ms}{MS}{mass spectrometry}

\newacronym{nbm}{NBM}{neutral beam monitor}
\newacronym{neb}{NEB}{nudged elastic band}
\newacronym{nim}{NIM}{nuclear instrumentation module}
\newacronym{nmr}{NMR}{nuclear magnetic resonance}
\newacronym{no}{NO}{nuclear orientation}
\newacronym{nqr}{NQR}{nuclear quadrupole resonance}
\newacronym{nrc}{NRC}{National Research Council of Canada}
\newacronym{nserc}{NSERC}{Natural Sciences and Engineering Research Council of Canada}

\newacronym{oa}{OA}{optical absorption}

\newacronym{pac}{PAC}{perturbed angular correlation}
\newacronym{pad}{PAD}{perturbed angular distribution}
\newacronym{pas}{PAS}{principle axis system}
\newacronym{pchip}{PCHIP}{piecewise cubic Hermite interpolating polynomial}
\newacronym{pdf}{PDF}{probability density function}
\newacronym{pld}{PLD}{pulsed laser deposition}
\newacronym{ppms}{PPMS}{physical property measurement system}
\newacronym{psi}{PSI}{Paul Scherrer Institute}

\newacronym{qens}{QENS}{quasielastic neutron scattering}
\newacronym{ql}{QL}{quintuple layer}
\newacronym{qo}{QO}{quantum oscillations}

\newacronym{rbs}{RBS}{Rutherford backscattering}
\newacronym{rf}{RF}{radio frequency}
\newacronym{rheed}{RHEED}{reflection high-energy electron diffraction}
\newacronym{rib}{RIB}{radioactive ion beam}
\newacronym{rkky}{RKKY}{Ruderman–Kittel–Kasuya–Yosida}
\newacronym{rrr}{RRR}{residual-resistivity ratio}
\newacronym{rtil}{RTIL}{room temperature ionic liquid}

\newacronym{sae}{SAE}{spin-alignment echo}
\newacronym{sans}{SANS}{small angle neutron scattering}
\newacronym{si}{SI}{International System of Units}
\newacronym{sims}{SIMS}{secondary ion mass spectrometry}
\newacronym{slr}{SLR}{spin-lattice relaxation}
\newacronym{sms}{S\ensuremath{\mu}S}{Swiss Muon Source}
\newacronym[sort={S/N}]{snr}{\textit{S}/\textit{N}}{signal-to-noise ratio}
\newacronym{squid}{SQUID}{superconducting quantum interference device}
\newacronym{srf}{SRF}{superconducting radio frequency}
\newacronym{srim}{SRIM}{Stopping and Range of Ions in Matter}
\newacronym{ssid}{SSID}{solid-state ionic device}
\newacronym{ssr}{SSR}{spin-spin relaxation}
\newacronym{stm}{STM}{scanning tunnelling microscopy}
\newacronym{sts}{STS}{scanning tunnelling spectroscopy}

\newacronym{ti}{TI}{topological insulator}
\newacronym{tof}{ToF}{time-of-flight}
\newacronym{trim}{TRIM}{Transport and Range of Ions in Matter}
\newacronym{tss}{TSS}{topological surface state}
\newacronym{tmd}{TMD}{transition metal dichalcogenide}

\newacronym{uhv}{UHV}{ultra-high vacuum}

\newacronym{vdw}{vdW}{van der Waals}
\newacronym{vft}{VFT}{Vogel-Fulcher-Tammann}

\newacronym{xrd}{XRD}{x-ray diffraction}
\newacronym{xrr}{XRR}{x-ray reflection}

\newacronym{ybco}{YBCO}{\ch{YBa2Cu3O_{6+x}}}
\newacronym{ysz}{YSZ}{yttria-stabilized zirconia}

\newacronym[sort={muSR}]{musr}{\ensuremath{\mu}SR}{muon spin spectroscopy}
\newacronym{alc-musr}{ALC-\ensuremath{\mu}SR}{avoided level crossing muon spin rotation}
\newacronym{le-musr}{LE-\ensuremath{\mu}SR}{low-energy \ensuremath{\mu}SR}
\newacronym{le-musr2}{LE-\ensuremath{\mu}SR}{low-energy muon spin spectroscopy}

\newacronym{lf-musr}{LF-\ensuremath{\mu}SR}{longitudinal field muon spin rotation}
\newacronym{rf-musr}{RF-\ensuremath{\mu}SR}{radio frequency muon spin rotation}
\newacronym{tf-musr}{TF-\ensuremath{\mu}SR}{transverse field muon spin rotation}
\newacronym{zf-musr}{ZF-\ensuremath{\mu}SR}{zero field muon spin rotation}

\begin{document}

\title{
	Implantation studies of low-energy positive muons in niobium thin films
}

\author{Ryan~M.~L.~McFadden}
\email[E-mail: ]{rmlm@triumf.ca}
\affiliation{TRIUMF, 4004 Wesbrook Mall, Vancouver, BC V6T~2A3, Canada}
\affiliation{Department of Physics and Astronomy, University of Victoria, 3800 Finnerty Road, Victoria, BC V8P~5C2, Canada}

\author{Andreas~Suter}
\email[E-mail: ]{andreas.suter@psi.ch}
\affiliation{PSI Center for Neutron and Muon Sciences CNM, Forschungsstrasse 111, 5232 Villigen PSI, Switzerland}

\author{Leon~Ruf}
\affiliation{Department of Physics, University of Konstanz, Universitaetsstrasse 10, 78464 Konstanz, Germany}

\author{Angelo~Di~Bernardo}
\affiliation{Department of Physics, University of Konstanz, Universitaetsstrasse 10, 78464 Konstanz, Germany}
\affiliation{Department of Physics, University of Salerno, Via Giovanni Paolo II 132, 84084 Fisciano SA, Italy}

\author{Arnold~M.~Müller}
\affiliation{Laboratory of Ion Beam Physics, ETH Zurich, HPK H32 Otto-Stern-Weg 5, 8093 Zurich, Switzerland}

\author{Thomas~Prokscha}
\affiliation{PSI Center for Neutron and Muon Sciences CNM, Forschungsstrasse 111, 5232 Villigen PSI, Switzerland}

\author{Zaher~Salman}
\affiliation{PSI Center for Neutron and Muon Sciences CNM, Forschungsstrasse 111, 5232 Villigen PSI, Switzerland}

\author{Tobias~Junginger}
\affiliation{TRIUMF, 4004 Wesbrook Mall, Vancouver, BC V6T~2A3, Canada}
\affiliation{Department of Physics and Astronomy, University of Victoria, 3800 Finnerty Road, Victoria, BC V8P~5C2, Canada}


\date{\today}

\begin{abstract}
Here we study the range of \unit{\kilo\electronvolt} positive muons $\mu^{+}$ implanted in
\ch{Nb2O5}($x$~\unit{\nano\meter})/\ch{Nb}($y$~\unit{\nano\meter})/\ch{SiO2}(\qty{300}{\nano\meter})/Si
[$x = \qty{3.6}{\nano\meter}, \qty{3.3}{\nano\meter}$; $y = \qty{42.0}{\nano\meter}, \qty{60.1}{\nano\meter}$]
thin films using \gls{le-musr2}.
At implantation energies $\qty{1.3}{\kilo\electronvolt} \leq E \leq \qty{23.3}{\kilo\electronvolt}$,
we compare the measured diamagnetic $\mu^{+}$ signal fraction $f_{\mathrm{dia.}}$ against
predictions derived from implantation profile simulations using the
\texttt{TRIM.SP} Monte Carlo code.
Treating the implanted $\mu^{+}$ as light protons $p^{+}$,
we find that simulations making use of
updated stopping cross section data
are in good agreement with the \gls{le-musr2} measurements,
in contrast to parameterizations found in earlier tabulations.
Implications for other studies relying on accurate
$\mu^{+}$ stopping information are discussed.
\end{abstract}

\maketitle
\glsresetall

\section{
	Introduction
	\label{sec:introduction}
}

The positive muon $\mu^{+}$
(spin $S_{\mu} = 1/2$;
gyromagnetic ratio $\gamma_{\mu} / (2 \pi) = \qty{135.538 809 4 \pm 0.000 003 0}{\mega\hertz\per\tesla}$~\cite{2021-Tiesinga-RMP-93-025010};
mean lifetime $\tau_{\mu} = \qty{2.1969811 \pm 0.0000022}{\micro\second}$~\cite{2022-Workman-PTEP-2022-083C01-short};
mass $m_{\mu} = \qty{0.113 428 925 9 \pm 0.000 000 002 5}{u}$~\cite{2021-Tiesinga-RMP-93-025010})
finds widespread use in the study of condensed matter,
with the leptonic elementary particle
serving as \gls{musr}['s] sensitive ``spin probe''~\cite{2022-Hillier-NRMP-2-4}.
In \gls{musr},
implanted $\mu^{+}$ are used to monitor the local electromagnetic environment
in a target of interest,
analogous to ``conventional'' \gls{nmr} measurements using stable nuclei
(see, e.g.,~\cite{2019-Pell-PNMRS-111-1}).
Naturally,
the range of applications is quite broad,
and
with over half-a-century of development and refinement~\cite{2012-Brewer-PP-30-2},
\gls{musr} is now routinely applied to challenging problems~\footnote{That is, physical systems that are either prohibitively difficult or impossible to study through more ``conventional'' means of interrogation.}
in physics, chemistry, and materials science
(see, e.g.,~\cite{2011-Yaouanc-MSR,2021-Blundell-MSI,2024-Amato-IMSS,2024-Fleming-MSSMACMS}).

In the last few decades,
the capabilities of \gls{musr} have been expanded to the realm of nanoscience
through the development of a contemporary variant called \gls{le-musr}~\cite{2004-Bakule-CP-45-203,2004-Morenzoni-JPCM-16-S4583}.
\Gls{le-musr}['s] key feature is the use of $\mu^{+}$ beams with precisely controlled implantation
energies $E \lesssim \qty{30}{\kilo\electronvolt}$,
which imparts the variant with \emph{spatial-resolution}
over subsurface depths on the order of
10s to 100s of nanometers~\cite{2002-Morenzoni-NIMB-192-245}.
With the exception of a few complementary~\cite{2000-Kiefl-PB-289-640} techniques
(e.g., $\beta$-radiation-detected \gls{nmr}~\cite{2015-MacFarlane-SSNMR-68-1,2022-MacFarlane-ZPC-236-757}),
\gls{le-musr} offers a unique combination of electromagnetic
and
spatial sensitivity,
making it ideally suited for the study of exotic (sub)surface phenomenon
(see, e.g.,~\cite{1999-Niedermayer-PRL-93-3932,2004-Suter-PRL-92-087001,2015-DiBernardo-PRX-5-041021,2016-Flokstra-NP-12-57,2019-Meier-PRX-9-011011,2020-Krieger-PRL-125-026802,2021-Fittipaldi-NC-12-5792,2021-Alpern-PRM-5-114801,2022-Fowlie-NP-18-1043}).
Though quite diverse in scope,
these studies all share a common trait:
they make explicit use of $\mu^{+}$'s $E$-dependent stopping distribution
$\rho(z, E)$ to extract spatial information from the analysis.

Generally,
information on $\rho(z, E)$ is not derived directly from a \gls{le-musr} measurement
(cf.~\cite{2004-Morenzoni-JPCM-16-S4583}),
meaning it must be obtained through alternative means.
The most common approach is through separate simulations of the implantation process,
with Monte Carlo \gls{bca} codes
(e.g., \texttt{SRIM}~\cite{srim}
or
\texttt{TRIM.SP}~\cite{1984-Biersack-APA-34-73,1991-Eckstein-SSMS-10,1994-Eckstein-REDS-1-239})
being widely used for this purpose~\footnote{Note that, as discussed in Ref.~\cite{2002-Morenzoni-NIMB-192-245}, the use of \texttt{SRIM}~\cite{srim} should be avoided for simulating the implantation of $\mu^{+}$ at \unit{\kilo\electronvolt} energies.}.
As the interaction of $\mu^{+}$ with matter follows that of
a \emph{very} light proton $p^{+}$
($m_{\mu^{+}} / m_{p^{+}} \approx 1 / 9$),
it is possible to leverage what is known about $p^{+}$
implantation~\cite{1977-Anderson-SRIM-3,1993-ICRU-49,1991-Eckstein-SSMS-10,srim}
and
adapt such codes for $\mu^{+}$ implantation~\cite{2002-Morenzoni-NIMB-192-245}.
While such simulations have been shown to accurately reproduce the $\mu^{+}$ range in
metals such as \ch{Al}, \ch{Cu}, \ch{Ag}, and \ch{Au}~\cite{2000-Gluckler-PB-289-658,2002-Morenzoni-NIMB-192-245},
similar checks in other elements remain lacking
---
particularly those where experimental stopping data is either sparse
(e.g., \ch{Na}, \ch{Ru}, or \ch{Eu})
or
absent
(e.g., \ch{Ho}, \ch{Rh}, or \ch{Pr})~\cite{2017-Montanari-NIMB-408-50,2024-Montanari-NIMB-551-165336}.
Though this lack of data can be mitigated to some extent through contemporary predictive
techniques
(see, e.g.,~\cite{2020-Parfitt-NIMB-478-21,2022-BivortHaiek-JAP-132-245103,2023-Akbari-NIMB-538-8,2024-Minagawa-NIMB-553-165383}),
\emph{experimental verification} remains the best means of achieving the accuracy desired for depth-resolved
\gls{le-musr} measurements.

To illustrate the impact of accurate stopping profiles,
we consider the elemental metal \ch{Nb}.
Despite being one of the most widely studied superconductors,
interest in interrogating its physical properties
(especially using implanted spin probes~\cite{2024-Junginger-FEM-4-1346235})
remains strong,
largely due to the element's importance for \gls{srf} technologies~\cite{2023-Padamsee-SRTA}.
\Gls{le-musr} has contributed to understanding the element's intrinsic Meissner response~\cite{2005-Suter-PRB-72-024506,McFadden-tbp},
along with how it is modified through different surface
treatments~\cite{2014-Romanenko-APL-104-072601,2023-McFadden-PRA-19-044018,2024-McFadden-APL-124-086101,2024-McFadden-AIPA-14-095320}
or
coatings~\cite{2024-Asaduzzaman-SST-37-025002}.
For this purpose,
accurate knowledge of $\rho(z,E)$ is crucial,
as it dictates the absolute value of the superconducting length scales
derived from the measurements.
It is apparent that early \gls{le-musr} measurements 
(see, e.g.,~\cite{2005-Suter-PRB-72-024506,2014-Romanenko-APL-104-072601})
find \ch{Nb}'s London penetration depth $\lambda_{\mathrm{L}}$ to be
considerably shorter than older, widely cited estimates
(see, e.g.,~\cite{1965-Maxfield-PR-139-A1515}),
suggesting the discrepancy may be \emph{systematic} in origin.

Recently,
it was suggested in Ref.~\cite{2023-McFadden-PRA-19-044018}
that refinements to the longstanding tabulations of
Varelas-Biersack~\cite{1970-Varelas-NIM-79-213} coefficients describing
$p^{+}$ stopping in \ch{Nb}~\cite{1977-Anderson-SRIM-3,1993-ICRU-49},
which are used by \texttt{TRIM.SP}~\cite{1984-Biersack-APA-34-73,1991-Eckstein-SSMS-10,1994-Eckstein-REDS-1-239},
were possible
through evaluation of the \gls{iaea}['s] database on electronic stopping
powers~\cite{2017-Montanari-NIMB-408-50,2024-Montanari-NIMB-551-165336}.
Specifically,
using all available stopping data~\cite{1984-Sirotinin-NIMB-4-337,1986-Bauer-NIMB-13-201,1988-Ogino-NIMB-33-155,2020-Moro-PRA-102-022808}
(apart from a dataset of clear outliers~\cite{1973-Behr-TSF-19-247}),
a new Varelas-Biersack~\cite{1970-Varelas-NIM-79-213}
parameterization was proposed~\cite{2023-McFadden-PRA-19-044018},
yielding magnetic penetration depths derived from \gls{le-musr}
consistent with \ch{Nb}'s clean-limit value \qty{\sim 29}{\nano\meter}
(see, e.g.,~\cite{2023-McFadden-PRA-19-044018,2023-McFadden-JAP-134-163902}).
While this agreement implies the correctness of the refined $\rho(z,E)$
simulations,
no independent check was performed.

\begin{figure}
	\centering
	\includegraphics[width=1.0\columnwidth]{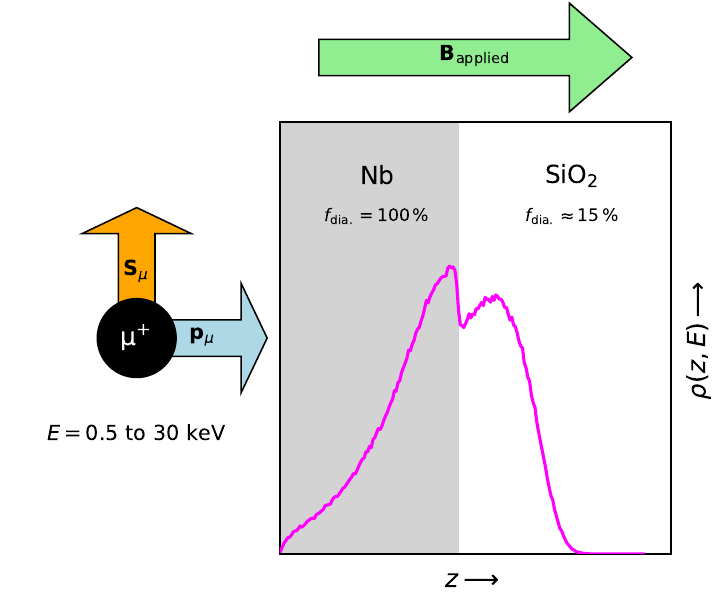}
	\caption{
		\label{fig:experiment-sketch}
		Sketch of the $\mu^{+}$ implantation experiment using \gls{le-musr}.
		Muons with their spin direction $\mathbf{S}_{\mu}$ perpendicular to their
		momentum $\mathbf{p}_{\mu}$ are implanted in \ch{Nb}/\ch{SiO2} films
		at energies $E \in [0.5, 30]$~\unit{\kilo\electronvolt},
		such that their stopping profile $\rho(z,E)$ overlaps with the two
		material layers.
		An external magnetic field
		$\mathbf{B}_{\mathrm{applied}} \approx \qty{10}{\milli\tesla} \parallel \hat{\mathbf{z}}$
		is applied perpendicular to $\mathbf{S}_{\mu}$,
		such that only the fraction of $\mu^{+}$ stopped in a diamagnetic environment $f_{\mathrm{dia.}}$
		is observable.
		As $f_{\mathrm{dia.}}$ deviates significantly between the two layers
		(see the sketch's inset),
		the \gls{le-musr} signal amplitude is expected to decrease as more $\mu^{+}$ stop
		in the \ch{SiO2} layer.
	}
\end{figure}

In this work,
we expand upon the $\rho(z, E)$ refinements suggested in Ref.~\cite{2023-McFadden-PRA-19-044018}
and
study the range of $\mu^{+}$ in \ch{Nb} using \gls{le-musr}.
Following the approach in Refs.~\cite{2000-Gluckler-PB-289-658,2002-Morenzoni-NIMB-192-245},
we implant $\mu^{+}$ in \ch{Nb} films deposited on amorphous \ch{SiO2} for a span of energies $E$
that result in $\rho(z,E)$s which overlap the two layers.
To determine the $\mu^{+}$ stopping environment,
we exploit the well-known property that,
upon thermalization,
all implanted $\mu^{+}$ end up in a diamagnetic environment in metals,
characterized by a Larmor frequency:
\begin{equation}
	\label{eq:larmor}
	\omega_{\mu} = \gamma_{\mu} B ,
\end{equation}
where $B$ is the local magnetic flux density at the $\mu^{+}$ stopping site.
By contrast,
in the wide-gap insulator \ch{SiO2}
the majority of $\mu^{+}$ bind with a ``free'' electron $e^{-}$
(e.g., those liberated during the implantation process)
to form the paramagnetic
atomic-like state muonium
($\ch{Mu} \equiv \mu^{+} + e^{-}$),
whose much larger gyromagnetic ratio $\gamma_{\ch{Mu}} \approx 103 \gamma_{\mu}$
makes it spectroscopically distinct from diamagnetic $\mu^{+}$.
In modest magnetic fields $B \gtrsim \qty{10}{\milli\tesla}$,
spin-precession of the paramagnetic fraction is fast enough that,
in the absence of specialized instrumentation providing high time-resolution,
it manifests as a \emph{loss} of signal,
with only the small diamagnetic fraction
$f_{\mathrm{dia.}} \approx \qty{15}{\percent}$~\cite{1984-Spencer-HI-18-567,2002-Morenzoni-NIMB-192-245,2003-Prokscha-PB-326-51,2007-Prokscha-PRL-98-227401,2012-Antognini-PRL-108-143401}
in \ch{SiO2} remaining observable.
A sketch of the experiment is given in \Cref{fig:experiment-sketch}.
By comparing the evolution of the \gls{le-musr} signal amplitude vs.\ $E$
in several \ch{Nb}/\ch{SiO2} films
against results from $\mu^{+}$ implantation simulations,
we find that a refined stopping power parameterization
is necessary to
accurately reproduce the $\mu^{+}$ range in \ch{Nb}.

\section{
	Experiment
	\label{sec:experiment}
}

\subsection{
	Samples
	\label{sec:experiment:samples}
}

In the present study,
we make use of two \ch{Nb} films:
one thin
and
another thicker
sample,
each deposited on
\ch{SiO2}(\qty{300}{\nano\meter})/\ch{Si} ($n$-doped, \qty{\sim 50}{\ohm\centi\meter})
substrates
(lateral dimensions of
\qtyproduct{\sim 2.5 x 2.5}{\centi\meter}
and
\qtyproduct{\sim 3 x 2}{\centi\meter}
for the thinner and thicker samples, respectively).
Both films were grown in an \gls{uhv} chamber using a combination of \gls{dc}
and \gls{rf} magnetron sputtering from three \ch{Nb} targets
(purity \qty{> 99.995}{\percent}).
The chamber's base (residual) pressure before growth was lower than
\qty{5.0e-9}{\torr}
and
an \ch{Ar} pressure of \qty{3}{\milli\torr} was used during deposition of the \ch{Nb} films
with a power of \qty{250}{\watt} applied to each \ch{Nb} target.
These settings yielded a deposition rate of \qty{\sim 0.4}{\nano\meter\per\second}.
Following synthesis,
film thickness were confirmed independently by \gls{xrr} and \gls{rbs} measurements
(see \Cref{sec:thickness}),
revealing slightly thinner dimensions for the \ch{Nb} layer
and 
the presence of a thin surface niobium pentoxide
[\ch{Nb2O5}(\qty{3.3}{\nano\meter})/\ch{Nb}(\qty{42.0}{\nano\meter})/\ch{SiO2}
and
\ch{Nb2O5}(\qty{3.6}{\nano\meter})/\ch{Nb}(\qty{60.1}{\nano\meter})/\ch{SiO2}
for the thinner and thicker films,
respectively].

\subsection{
	\acrshort{le-musr} Measurements
	\label{sec:experiment:le-musr}
}

\Gls{le-musr} measurements were performed at the \gls{sms} in the \gls{psi},
located in Villigen, Switzerland.
Using the $\mu E4$ beamline~\cite{2008-Prokscha-NIMA-595-317},
a \qty{\sim 100}{\percent} spin-polarized $\mu^{+}$ beam
(intensity of \qty{\sim e4}{\per\second})
was generated by moderating a \qty{\sim 4}{\mega\electronvolt}  ``surface'' $\mu^{+}$ beam
using a film of condensed cryogenic gas~\cite{1994-Morenzoni-PRL-72-2793,2001-Prokscha-ASS-172-235}
and electrostatically re-accelerating the eluting epithermal
(\qty{\sim 15}{\electronvolt})
muons to 
\qty{\sim 10}{\kilo\electronvolt} or \qty{\sim 15}{\kilo\electronvolt}.
The beam was delivered to a dedicated spectrometer~\cite{2000-Morenzoni-PB-289-653,2008-Prokscha-NIMA-595-317,2012-Salman-PP-30-55}
using electrostatic optics housed within an \gls{uhv} beamline,
with the $\mu^{+}$ arrival times triggered on a thin
(\qty{\sim 8}{\nano\meter}) carbon foil detector.
Note that passage through the foil results in both a slight reduction in
the beam's mean kinetic energy (\qty{\sim 1}{\kilo\electronvolt})
and
introduces a small (asymmetric) energy spread (\qty{\sim 450}{\electronvolt}).
Control over the $\mu^{+}$ implantation energy
(and the $\mu^{+}$ stopping depth)
is achieved by biasing
an electrically isolated sample holder using a \gls{hv} power supply.
The beam was implanted into the \ch{Nb}/\ch{SiO2} films
affixed to the spectrometer's cryostat,
which was maintained at \qty{200}{\kelvin} during the measurements
(i.e., to mitigate the accumulation of water atop the film's surface
at the sample chamber's pressure \qty{< e-7}{\milli\bar}).

In \gls{le-musr},
the temporal evolution of the $\beta$-decay \emph{asymmetry} $\mathcal{A}(t)$,
is monitored using a set of decay positron detectors $i$ surrounding the sample environment.
The measured $\mathcal{A}$ is proportional to the spin-polarization of the $\mu^{+}$ ensemble:
\begin{equation*}
	P_{\mu}(t) \equiv \langle S_{\mu, \perp} \rangle / S_{\mu},
\end{equation*}
where $\langle S_{\mu, \perp} \rangle / S_{\mu} \in [-1, 1]$ is the (normalized) expectation value
transverse to the applied field.
The count rate in a single detector $N_{i}$ is given by:
\begin{equation}
	\label{eq:counts}
	N_{i}(t) = N_{0, i} \exp \left ( - \frac{t}{\tau_{\mu}} \right ) \left [ 1 + \mathcal{A}_{i}(t) \right ] + b_{i} ,
\end{equation}
where $N_{0, i}$ and $b_{i}$ are the incoming rates of
``good'' and ``background''
decay events~\footnote{More precisely, $N_{0,i}$ is proportional to the total number of muons implanted (i.e., the size of the spin ensemble) and $b_{i}$ is the (virtually) time-independent count rate from muon-positron decays that are uncorrelated (see, e.g,~\cite{2024-Amato-IMSS}). Typically, $b_{i}$ is \qty{\sim 1}{\percent} of $N_{0, i}$.},
and
\begin{equation}
	\label{eq:asymmetry}
	\mathcal{A}_{i}(t) = \mathcal{A}_{0, i} P_{\mu}(t) ,
\end{equation}
where $\mathcal{A}_{0, i}$ is a proportionality constant typically $\lesssim 0.3$.
Note that,
written in this manner,
\Cref{eq:counts,eq:asymmetry} implicitly account for any ``instrumental'' differences
between individual detectors,
with the $P_{\mu}(t)$ treated as common among them.

Following implantation,
the $\mu^{+}$ spins reorient in their local magnetic environment
(i.e., at their stopping site).
When the local field is transverse to the ensemble's spin direction,
$P_{\mu}(t)$ will precess at a rate equal to the probe's Larmor frequency $\omega_{\mu}$,
In the experiments performed here,
a so-called transverse-field geometry was used
(see, e.g.,~\cite{2004-Bakule-CP-45-203,2011-Yaouanc-MSR}),
wherein an external field $B_\mathrm{applied} \approx \qty{10}{\milli\tesla}$
is applied perpendicular to $P_{\mu}(t)$'s initial direction
(see \Cref{fig:experiment-sketch}).
In this configuration,
as mentioned in \Cref{sec:introduction},
only $\mu^{+}$ in diamagnetic environments contribute to the observable signal,
with the paramagnetic ``\ch{Mu}'' fraction de-phasing before the earliest time
bins~\footnote{At lower applied fields, it is possible to resolve the approximately degenerate \ch{Mu} spin-precession frequencies in high-statistics measurements (cf.~\cite{2002-Morenzoni-NIMB-192-245,2003-Prokscha-PB-326-51,2007-Prokscha-PRL-98-227401}).}.
Consequently,
at the temperature of the present measurements,
the temporal evolution of spin-polarization follows~\footnote{Note that the field distribution in \ch{Nb} is intrinsically Gaussian, but ``narrows'' to a Lorentzian under modest muon motion and gives rise to the exponential term in \Cref{eq:polarization} (see, e.g.,~\cite{2024-Amato-IMSS}). When the depolarization rate is slow compared to the inverse muon lifetime (as in the case here), the form of the relaxation ``envelope'' for the two distribution types is nearly indistinguishable. Empirically, we find a slightly better fit is obtained if exponential depolarization is assumed.}:
\begin{equation}
	\label{eq:polarization}
	P_{\mu}(t) \approx \exp \left ( - \lambda t \right) \cos \left ( \omega_{\mu} t + \phi_{i} \right ) ,
\end{equation}
where $\lambda$ is the damping rate proportional to the width of the (Lorentzian) field distribution,
and
$\phi_{i}$ is a detector-dependent phase factor.

\subsection{
	Implantation Simulations
	\label{sec:experiment:trimsp}
}

\begin{figure}
	\centering
	\includegraphics[width=1.0\columnwidth]{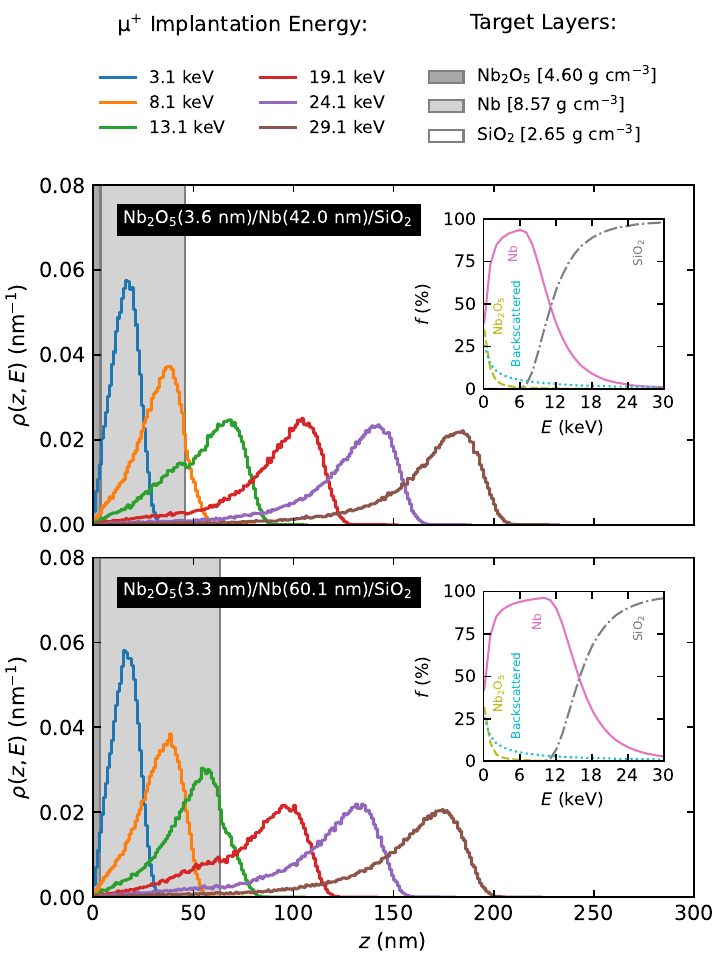}
	\caption{
		\label{fig:implantation-profiles}
		Typical stopping profiles $\rho(z, E)$ for $\mu^{+}$ implanted in
		\ch{Nb2O5}($x$~\unit{\nano\meter})/\ch{Nb}($y$~\unit{\nano\meter})/\ch{SiO2}
		[$x = \qty{3.6}{\nano\meter}, \qty{3.3}{\nano\meter}$; $y = \qty{42.0}{\nano\meter}, \qty{60.1}{\nano\meter}$]
		targets
		at different energies $E \leq \qty{30}{\kilo\electronvolt}$.
		The profiles were simulated using the \gls{bca} Monte Carlo code
		\texttt{TRIM.SP}~\cite{1984-Biersack-APA-34-73,1991-Eckstein-SSMS-10,1994-Eckstein-REDS-1-239}
		for \num{e5} projectiles,
		with the results represented as histograms with \qty{2}{\nano\meter} bins.
		Note that these simulations make use of our revised Varelas-Biersack~\cite{1970-Varelas-NIM-79-213}
		parameterization of each target atom's electronic stopping cross section
		(see \Cref{sec:cross-sections}).
		The evolution of the $\mu^{+}$ stopping fraction $f$ with $E$ is shown in
		each plot's inset. 
		Further simulation details are described in \Cref{sec:experiment:trimsp}
		and Refs.~\cite{2002-Morenzoni-NIMB-192-245,2023-McFadden-PRA-19-044018}.
	}
\end{figure}

Treating $\mu^{+}$ as a light $p^{+}$,
we simulate its implantation into our films using the \gls{bca} Monte Carlo code
\texttt{TRIM.SP}~\cite{1984-Biersack-APA-34-73,1991-Eckstein-SSMS-10,1994-Eckstein-REDS-1-239}.
The simulation inputs
(e.g., projectile energy spread, projectile-target atom interaction potential, etc.)
were tuned for \gls{psi}['s] \gls{le-musr} setup 
(see \Cref{sec:experiment:le-musr}),
with a thorough account of these details given
elsewhere~\cite{2002-Morenzoni-NIMB-192-245,2023-McFadden-PRA-19-044018}.
Crucial to accurate $\mu^{+}$ range estimation,
these simulations make use of the empirical Varelas-Biersack~\cite{1970-Varelas-NIM-79-213}
parameterization of electronic stopping cross sections $\tilde{S}_{e}$,
with values for different (elemental) targets tabulated in the
literature~\cite{1977-Anderson-SRIM-3,1993-ICRU-49}.
As part of this work,
we consider the validity of these compilations using stopping power data
available from the \gls{iaea}['s] database~\cite{2017-Montanari-NIMB-408-50,2024-Montanari-NIMB-551-165336}
(see \Cref{sec:cross-sections}).
From new Varelas-Biersack~\cite{1970-Varelas-NIM-79-213}
fits to the $\tilde{S}_{e}$ data,
we determine improved parameterizations for all target elements considered here
(i.e., \ch{O}, \ch{Si}, and \ch{Nb}),
with the most consequential updates being for \ch{Nb}
(see Ref.~\cite{2023-McFadden-PRA-19-044018}).
Further details are given in \Cref{sec:cross-sections}.
Using these values,
along with those in the older tabulations~\cite{1977-Anderson-SRIM-3,1993-ICRU-49}
for comparison,
we simulated $\mu^{+}$ implantation into the \ch{Nb}/\ch{SiO2}/\ch{Si} films at different $E$,
using \num{e5} projectiles in each case.
Typical stopping profiles are shown in \Cref{fig:implantation-profiles}.

\section{
	Results
	\label{sec:results}
}

\begin{figure}
	\centering
	\includegraphics[width=1.0\columnwidth]{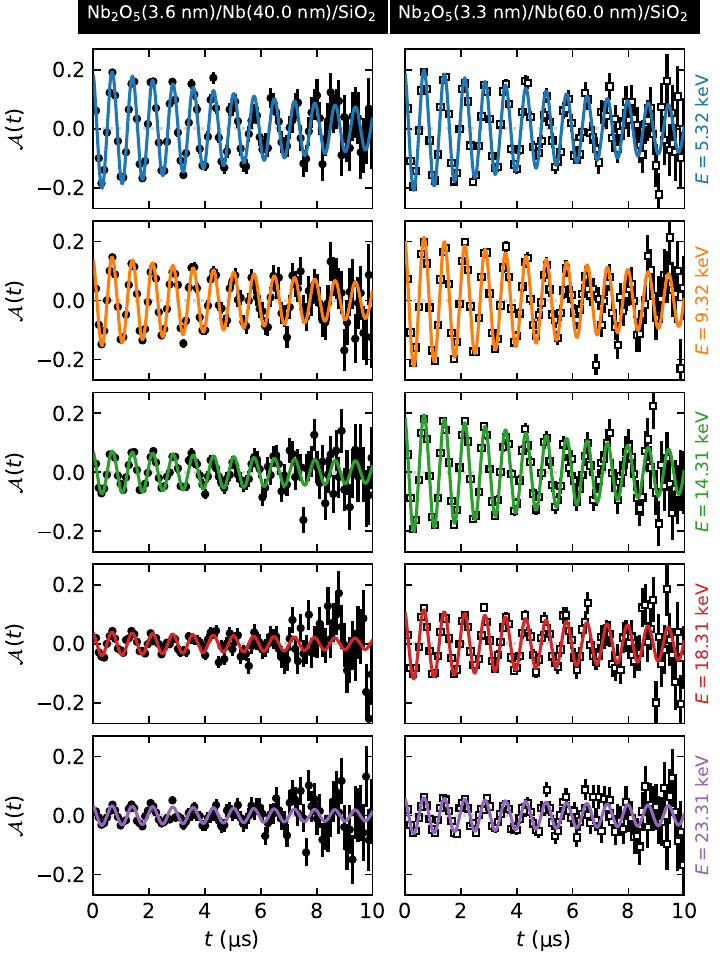}
	\caption{
		\label{fig:asymmetry}
		Typical time-differential \gls{le-musr} data in the
		\ch{Nb2O5}($x$~\unit{\nano\meter})/\ch{Nb}($y$~\unit{\nano\meter})/\ch{SiO2}
		[$x = \qty{3.6}{\nano\meter}, \qty{3.3}{\nano\meter}$; $y = \qty{42.0}{\nano\meter}, \qty{60.1}{\nano\meter}$]
		films measured at $T = \qty{200}{\kelvin}$ in a
		$B_{\mathrm{applied}} = \qty{10}{\milli\tesla}$ transverse-field
		at different implantation energies $E$
		and
		a $\mu^{+}$ transport bias $\mathrm{Tr} = \qty{15}{\kilo\electronvolt}$.
		Note that only the signal in one of four detectors
		(binned by a factor of \num{500})
		is shown for clarity. 
		The solid colored lines denote fits to \Cref{eq:counts,eq:asymmetry,eq:polarization},
		in excellent agreement with the data.
		The amplitude $\mathcal{A}(t=\qty{0}{\micro\second}) \equiv \mathcal{A}_{0}$ of the observable
		single is proportional to the population of implanted $\mu^{+}$ in a
		diamagnetic state,
		corresponding (predominantly) to $\mu^{+}$ stopped in the \ch{Nb} layer,
		which decreases with increasing $E$. 
	}
\end{figure}

\begin{figure}
	\centering
	\includegraphics[width=1.0\columnwidth]{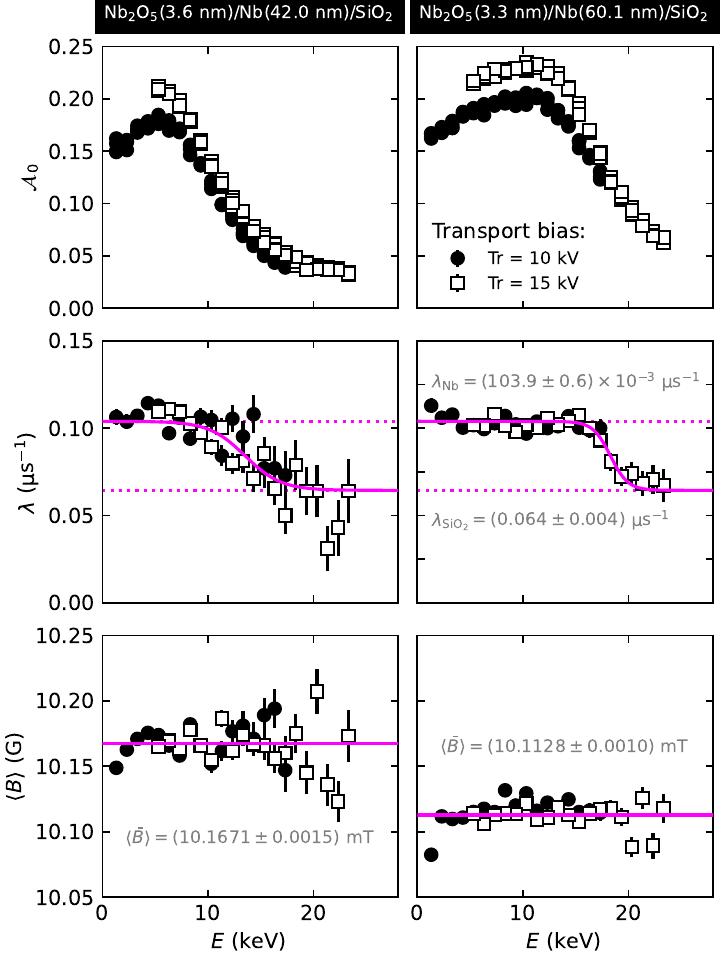}
	\caption{
		\label{fig:fitpar}
		Implantation energy $E$ dependence of the main
		fit parameters derived from the analysis of the \gls{le-musr}
		data in the
		\ch{Nb2O5}($x$~\unit{\nano\meter})/\ch{Nb}($y$~\unit{\nano\meter})/\ch{SiO2}
		[$x = \qty{3.6}{\nano\meter}, \qty{3.3}{\nano\meter}$; $y = \qty{42.0}{\nano\meter}, \qty{60.1}{\nano\meter}$]
		films at $T = \qty{200}{\kelvin}$ in a
		$B_{\mathrm{applied}} = \qty{10}{\milli\tesla}$ transverse-field
		using
		\Cref{eq:counts,eq:asymmetry,eq:polarization}.
		Here,
		$\mathcal{A}_{0}$ is the initial asymmetry,
		proportional to the population of $\mu^{+}$ stopped in a diamagnetic environment,
		$\lambda$ is the exponential damping rate,
		and
		$\langle B \rangle$ is the mean local field experienced by the spin-probes.
		While systematic differences are evident for the $\mathcal{A}_{0}$s determined
		at different transport biases $\mathrm{Tr}$
		(e.g., due to slightly different initial spin states),
		their $E$-dependence is identical up to a normalization factor.
		The sigmoidal $E$-dependence of $\lambda$ reflects the different damping rates
		for the \ch{Nb} and \ch{SiO2} layers,
		whose asymptotic values
		($\lambda_{\ch{Nb}}$ and $\lambda_{\ch{SiO2}}$)
		are given in the plot inset.
		Values for the $E$-independent local field values
		$\bar{\langle B \rangle}$ are also indicated.
	}
\end{figure}

Typical time-differential \gls{le-musr} data in the \ch{Nb}/\ch{SiO2} films
at different $E$ are shown in \Cref{fig:asymmetry}.
Coherent spin-precession is observed at all measurement conditions,
as evidenced by the weakly damped oscillations in $\mathcal{A}_{i}(t)$.
The most striking changes to the data is the loss in signal amplitude
$\mathcal{A}_{i}(t = \qty{0}{\micro\second}) \equiv A_{0,i}$ with increasing $E$,
corresponding to more $\mu^{+}$ stopping in \ch{SiO2} where
the diamagnetic fraction $f_{\mathrm{dia.}}$ is small
(see \Cref{fig:experiment-sketch,fig:implantation-profiles}).
The $E$ where the decrease in $\mathcal{A}_{0,i}$ onsets is different for the two films,
reflecting the differing thickness of their \ch{Nb} layers. 
To quantify these changes,
we fit the data to
\Cref{eq:larmor,eq:counts,eq:asymmetry,eq:polarization}
using \texttt{musrfit}~\cite{2012-Suter-PP-30-69},
yielding excellent agreement with the data
(typical reduced negative log-likelihood $\tilde{\mathcal{L}} \approx 1.01$ for each fit.).
The $E$-dependence of the main fit parameters extracted from this
procedure are shown in \Cref{fig:fitpar}
and
we consider them in detail below.

First we consider the $\mathcal{A}_{0,i}$s.
Mirroring the changes in signal amplitude with $E$ shown in \Cref{fig:asymmetry},
the $\mathcal{A}_{0,i}$s in both films decrease when $E \gtrsim \qty{10}{\kilo\electronvolt}$.
At lower $E$,
a film-dependent maximum is reached,
whereafter $\mathcal{A}_{0,i}$ decreases with decreasing $E$,
reflecting,
for example,
the increased probability of $\mu^{+}$ backscattering
off the sample target~\cite{2002-Morenzoni-NIMB-192-245,2023-Suter-JPCS-2462-012011}.
The intrinsic nature of these trends is confirmed by measurements
using a different $\mu^{+}$ transport bias $\mathrm{Tr}$,
which show the same $E$-dependence.
We note, however, that there is a small offset between the data acquired using
$\mathrm{Tr} = \qty{10}{\kilo\volt}$ and \qty{15}{\kilo\volt}.
This difference is \emph{instrumental},
reflecting,
for example,
slightly different initial polarizations of the $\mu^{+}$ beam
for the two $\mathrm{Tr}$ values,
and is easily rectified by a ``normalization'' factor on the order of unity.
Importantly,
this detail does not impact the extraction of $f_{\mathrm{dia.}}$ from either dataset
(see below).

\begin{figure*}
	\centering
	\includegraphics[width=1.0\textwidth]{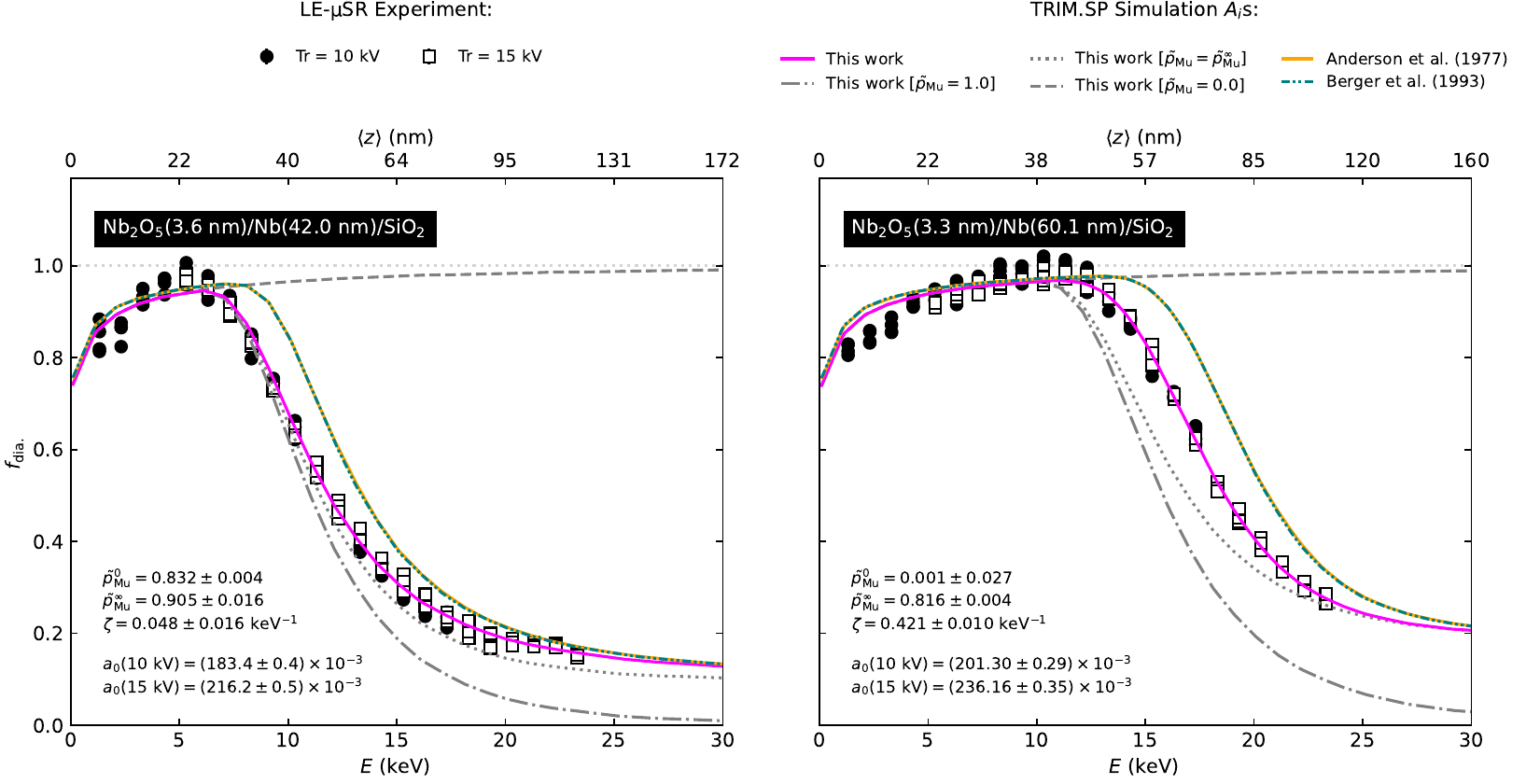}
	\caption{
		\label{fig:fractions}
		Measured diamagnetic fraction $f_{\mathrm{dia.}}$
		as a function of $\mu^{+}$ implantation energy $E$
		in the
		\ch{Nb2O5}($x$~\unit{\nano\meter})/\ch{Nb}($y$~\unit{\nano\meter})/\ch{SiO2}
		[$x = \qty{3.6}{\nano\meter}, \qty{3.3}{\nano\meter}$; $y = \qty{42.0}{\nano\meter}, \qty{60.1}{\nano\meter}$]
		thin films at $T = \qty{200}{\kelvin}$ in a transverse applied field
		$B_{\mathrm{applied}} \approx \qty{10}{\milli\tesla}$.
		Data points correspond to measurements by \gls{le-musr}
		using two different $\mu^{+}$ beam extraction biases $\mathrm{Tr}$.
		The
		lines indicate predictions by the \gls{bca} Monte Carlo code
		\texttt{TRIM.SP}~\cite{1984-Biersack-APA-34-73,1991-Eckstein-SSMS-10,1994-Eckstein-REDS-1-239}
		using different
		muonium formation probabilities $\tilde{p}_{\ch{Mu}}(E^{*})$
		and
		electronic stopping cross section parameterizations.
		Simulations using revised inputs
		derived from the \gls{iaea}['s] stopping database~\cite{2017-Montanari-NIMB-408-50,2024-Montanari-NIMB-551-165336}
		(extending the work in Ref.~\cite{2023-McFadden-PRA-19-044018})
		are in excellent agreement with the data,
		in contrast to predictions based on older cross section
		tabulations~\cite{1977-Anderson-SRIM-3,1993-ICRU-49},
		which systematically underestimate the range of $\mu^{+}$
		(i.e., predict larger $f_{\mathrm{dia.}}$s).
		Note that the $f_{\mathrm{dia.}}$s simulated using the older tabulations~\cite{1977-Anderson-SRIM-3,1993-ICRU-49}
		are nearly indistinguishable,
		owing to their similar stopping coefficients $A_{i}$ for \ch{O}, \ch{Si}, and \ch{Nb}
		(see \Cref{tab:coeff}). 
	}
\end{figure*}

Next we consider the depolarization rate $\lambda$,
whose main feature is a sigmoidal $E$-dependence,
gradually transitioning from a value of
\qty{\sim 0.11}{\per\micro\second}
to
\qty{\sim 0.07}{\per\micro\second}.
We interpret this change as reflecting differences
in the local field distribution inside the \ch{Nb}
and \ch{SiO2} layers,
with measurements at low (high) $E$s 
predominantly sampling the \ch{Nb} (\ch{SiO2}) layers.
Using a phenomenological model to describe this partitioning,
from its asymptotic limits we determine
$\lambda_{\ch{Nb}} = \qty{0.1039 \pm 0.0006}{\per\micro\second}$
and
$\lambda_{\ch{SiO2}} = \qty{0.064 \pm 0.004}{\per\micro\second}$.
Noting that the intrinsic local field distribution in \ch{Nb}
is Gaussian, characterized by a damping parameter
\qty{\sim 0.5}{\per\micro\second}
(see, e.g.,~\cite{1982-Boekema-PRB-26-2341,2013-Maisuradze-PRB-88-140509,2017-Junginger-SST-30-125013}),
the observed exponential damping and small $\lambda_{\ch{Nb}}$ indicate that $\mu^{+}$
is mobile at \qty{200}{\kelvin};
however,
its finite value suggests the absence of complete ``motional narrowing''
(cf.~\cite{1978-Birnbaum-PRB-17-4143}). 
While significant $\mu^{+}$ motion can complicate implantation studies,
site-to-site ``hopping'' rates $\nu \gtrsim \qty{e9}{\per\second}$
are required to appreciably affect $\mathcal{A}_{0,i}$~\cite{2002-Morenzoni-NIMB-192-245}.
We assert that the modest $\mu^{+}$ mobility at \qty{200}{\kelvin}~\cite{1978-Borghini-PRL-40-1723,1978-Birnbaum-PRB-17-4143}
is insufficient to be of detriment to our implantation studies.
We shall revisit this point later in \Cref{sec:discussion}.

Before proceeding,
we briefly touch upon the $B$s extracted from our fits to the \gls{le-musr} data.
Though some small scatter is evident
(e.g., at high $E$ where the signal is small),
consistent with expectations for a nonmagnetic heterostructure,
$B$ is found to be $E$-independent.
Note that the minor difference in the average values $\langle B \rangle$
determined for both samples
(see the inset in \Cref{fig:fitpar}) reflects their slightly different
geometries,
amounting to different demagnetization contributions to $B$
(see, e.g.,~\cite{2024-Amato-IMSS}).

With the main fit results outlined above,
we now evaluate the $E$-dependence of the $\mathcal{A}_{0,i}$s in \Cref{fig:fitpar}
and
convert them to $f_{\mathrm{dia.}}$s using results from the
\texttt{TRIM.SP}~\cite{1984-Biersack-APA-34-73,1991-Eckstein-SSMS-10,1994-Eckstein-REDS-1-239}
simulations of $\mu^{+}$ implantation
(described in \Cref{sec:experiment:trimsp}).
From the $\mu^{+}$ stopping fractions in \Cref{fig:implantation-profiles},
$f_{\mathrm{dia.}}$ may be calculated as~\cite{2002-Morenzoni-NIMB-192-245}:
\begin{equation}
	\label{eq:f_dia}
	f_{\mathrm{dia.}}(E) \approx f_{\ch{Nb2O5}}(E) + f_{\ch{Nb}}(E) + \left [ 1 - \tilde{p}_{\ch{Mu}}(E^{*}) \right ] f_{\ch{SiO2}}(E) , 
\end{equation}
where
$f_{i}(E) \in [0, 1]$ is the fraction of $\mu^{+}$ stopped in layer $i = \{ \ch{Nb2O5}, \ch{Nb}, \ch{SiO2} \}$
(the signal in \ch{Nb2O5} is assumed to be purely diamagnetic~\cite{arXiv:2312.10697}),
and
$\tilde{p}_{\ch{Mu}}(E^{*})$ is the \ch{Mu} formation probability in
\ch{SiO2}.
This latter quantity is known to be $E$-dependent,
decreasing from its ``bulk'' value as $E \rightarrow \qty{0}{\kilo\electronvolt}$~\cite{2002-Morenzoni-NIMB-192-245,2003-Prokscha-PB-326-51,2007-Prokscha-PRL-98-227401}.
We account for this detail explicitly using the mean energy $E^{*}$
of $\mu^{+}$ \emph{transmitted} through the \ch{Nb2O5}/\ch{Nb}
layers~\footnote{To determine $E^{*}$, separate \texttt{TRIM.SP} simulations were required where the \ch{SiO2} layer was omitted (see \Cref{sec:transmitted})~\cite{2000-Gluckler-PB-289-658,2002-Morenzoni-NIMB-192-245}.}:
\begin{equation}
	\label{eq:p_Mu}
	\tilde{p}_{\ch{Mu}} (E^{*}) \approx \left ( \tilde{p}_{\ch{Mu}}^{0} - \tilde{p}_{\ch{Mu}}^{\infty} \right ) \exp \left ( - \zeta E^{*} \right ) + \tilde{p}_{\ch{Mu}}^{\infty} ,
\end{equation}
where
$\tilde{p}_{\ch{Mu}}^{\infty}$ is the \ch{Mu} formation probability
in the high-$E^{*}$ limit,
$\tilde{p}_{\ch{Mu}}^{0}$ is the formation probability
for $E^{*} \rightarrow \qty{0}{\kilo\electronvolt}$,
and
$\zeta$ is a constant defining the interpolation between the two limits.
Note that while the form of \Cref{eq:p_Mu} is phenomenological,
it is consistent with the (limited) data on $E$-dependent \ch{Mu}
formation
(see, e.g.,~\cite{2002-Morenzoni-NIMB-192-245,2003-Prokscha-PB-326-51,2007-Prokscha-PRL-98-227401}).
The diamagnetic fraction is related to the observed $\mathcal{A}_{0}$s by:
\begin{equation}
	\label{eq:scale}
	\mathcal{A}_{0}(E, \mathrm{Tr}) = a_{0}(\mathrm{Tr}) f_{\mathrm{dia.}}(E) ,
\end{equation}
where $a_{0}(\mathrm{Tr})$ is a scaling factor that depends on transport bias
(cf.\ \Cref{fig:fitpar}).

To identify the $a_{0}(\mathrm{Tr})$s
and
the parameterization of $\tilde{p}_{\ch{Mu}}(E^{*})$,
for each sample
the $\mathcal{A}_{0,i}$s at each $\mathrm{Tr}$ were fit
with \Cref{eq:f_dia,eq:p_Mu,eq:scale}.
This was done for implantation simulations using our
revised Varelas-Biersack~\cite{1970-Varelas-NIM-79-213}
fit of the $\tilde{S}_{e}$ data for each element in our films
(see \Cref{sec:cross-sections}),
as well as values from older tabulations~\cite{1977-Anderson-SRIM-3,1993-ICRU-49}.
The fit results are shown in \Cref{fig:fractions},
with the resulting fit parameters given in the inset.
The best agreement is obtained for simulations using
our revised fit to the target element $\tilde{S}_{e}$s.
Interestingly,
$\tilde{p}_{\ch{Mu}}$'s $E$-dependence turned out to be somewhat different for the
two films;
however,
we note that both of their high-$E$ limiting values are
in good agreement with the range expected
for \unit{\kilo\electronvolt}
implantation energies~\cite{2002-Morenzoni-NIMB-192-245,2003-Prokscha-PB-326-51,2007-Prokscha-PRL-98-227401,2012-Antognini-PRL-108-143401}.
We shall consider this further in the subsequent section.

\section{
	Discussion
	\label{sec:discussion}
}

It is clear from \Cref{fig:fractions} that our revised $\tilde{S}_{e}$ coefficients
give the best agreement with the \gls{le-musr} data.
For $E \gtrsim \qty{5}{\kilo\electronvolt}$,
the $f_{\mathrm{dia.}}$ derived from the
\texttt{TRIM.SP}~\cite{1984-Biersack-APA-34-73,1991-Eckstein-SSMS-10,1994-Eckstein-REDS-1-239}
implantation simulations closely follows the \gls{le-musr} measurements,
accurately accounting for the partitioning of $\mu^{+}$ across the \ch{Nb} and \ch{SiO2}
layers.
In contrast,
the simulations relying on older $\tilde{S}_{e}$
tabulations~\cite{1977-Anderson-SRIM-3,1993-ICRU-49}
systematically overestimate $f_{\mathrm{dia.}}$
on the high $E$ side of its maximum.
This is (predominantly) a consequence of \ch{Nb}'s electronic stopping cross section
being overestimated near it's maximum in older tabulations~\cite{1977-Anderson-SRIM-3,1993-ICRU-49},
which was pointed out previously~\cite{2023-McFadden-PRA-19-044018}.
These measurements highlight the importance of having high-quality stopping data
publicly available~\cite{2017-Montanari-NIMB-408-50,2024-Montanari-NIMB-551-165336}
and
judiciously inspecting the details of the stopping profile simulations
used to complement a \gls{le-musr} experiment.

\begin{figure}
	\centering
	\includegraphics[width=1.0\columnwidth]{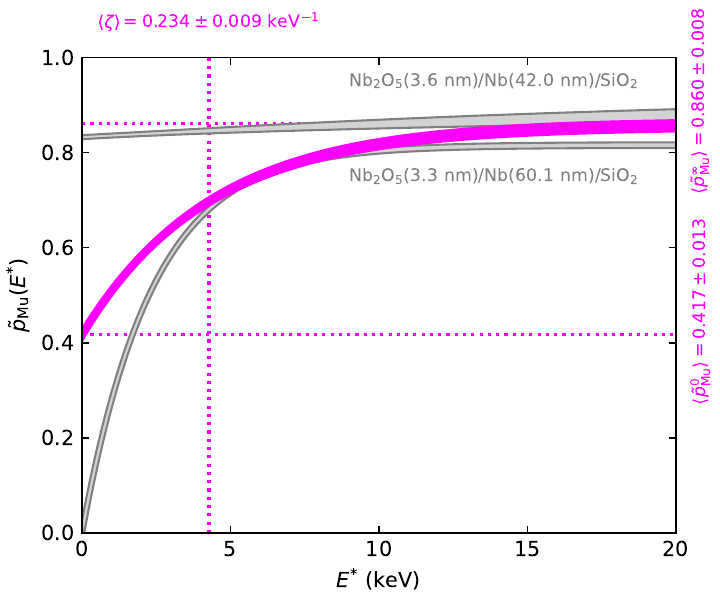}
	\caption{
		\label{fig:mu-formation}
		Muonium formation probability $\tilde{p}_{\ch{Mu}}$ as a function
		of the mean energy $E^{*}$ of $\mu^{+}$ transmitted through the
		\ch{Nb2O5}/\ch{Nb} layers.
		The shaded grey ``bands'' denote the \qty{\sim 68}{\percent} confidence region
		for each film's $\tilde{p}_{\ch{Mu}}(E^{*})$,
		as given by \Cref{eq:p_Mu} and the best-fit parameters listed in \Cref{fig:fractions}'s inset.
		The magenta ``band'' shows a similar confidence region for both film's average,
		which is characterized by the parameters in angular brackets $\langle \cdots \rangle$
		annotating the curve.
		The average behavior is in good accord with $\tilde{p}_{\ch{Mu}}$'s $E$-dependence
		in \ch{SiO2} samples~\cite{2002-Morenzoni-NIMB-192-245,2003-Prokscha-PB-326-51,2007-Prokscha-PRL-98-227401}.
	}
\end{figure}

As mentioned in \Cref{sec:results},
one of the ``degrees of freedom'' in our implantation profile modeling is the muonium
formation probability $\tilde{p}_{\ch{Mu}}(E^{*})$ in \ch{SiO2},
which warrants further inspection.
In ``bulk'' \gls{musr} where $\mu^{+}$ is implanted at \qty{\sim 4.1}{\mega\electronvolt},
$\tilde{p}_{\ch{Mu}}$ is close to unity,
with only a small \qty{\sim 15}{\percent} fraction of $\mu^{+}$ remaining
in a diamagnetic state in the insulating
oxide~\cite{1984-Spencer-HI-18-567,2002-Morenzoni-NIMB-192-245,2003-Prokscha-PB-326-51,2007-Prokscha-PRL-98-227401,2012-Antognini-PRL-108-143401}.
This remains constant until the $\mu^{+}$ implantation energy is lowered
to $\lesssim \qty{10}{\kilo\electronvolt}$,
whereafter $f_{\mathrm{dia.}}$ increases exponentially with decreasing $E$.
This behavior reflects the onset of \ch{Mu}'s delayed formation~\cite{2000-Brewer-PB-289-425}
once sufficient ``excess'' electrons \ch{e^{-}} become available
from $\mu^{+}$'s ionization track
(based on \ch{SiO2}'s electron-hole pair creation energy of \qty{\sim 18}{\electronvolt}~\cite{1975-Ausman-APL-26-173},
this corresponds to roughly \num{\sim 1000} \ch{e^{-}}~\cite{2007-Prokscha-PRL-98-227401}).
Qualitatively,
this is similar to what we observe in \Cref{fig:fractions},
where the important quantity is the energy $E^{*}$ of $\mu^{+}$ transmitted through the
\ch{Nb2O5}/\ch{Nb} layers
(see \Cref{sec:transmitted}),
which is akin to a reduced $E$.
While $\tilde{p}_{\ch{Mu}}$'s $E$-dependence is known to be sample-dependent
(see, e.g.,~\cite{2002-Morenzoni-NIMB-192-245,2003-Prokscha-PB-326-51,2007-Prokscha-PRL-98-227401}),
it is surprising that the $\tilde{p}_{\ch{Mu}}(E^{*})$ details differ for our two films,
particularly in the low-$E^{*}$ range
(see \Cref{fig:mu-formation}).
To explain this observation,
we consider the impact of $\mu^{+}$ \emph{energy straggling}
in our films.

Energy straggling is the variation in projectile energy 
following traversal through a target
(i.e., $E^{*} \pm \delta E^{*}$),
arrising from the stochastic nature of ion-implantation.
This effect is already integral to \gls{le-musr},
accounting for $E$'s \qty{\sim 450}{\electronvolt} distribution ``width''
from passage through a thin carbon-foil ``trigger''
prior to implantation~\footnote{This detail is accounted for in the muon-implantation simulations (see, e.g.,~\cite{2002-Morenzoni-NIMB-192-245,2023-McFadden-PRA-19-044018}).},
and
is similar to range straggling,
producing the spread in $\mu^{+}$'s stopping profiles
(see \Cref{fig:implantation-profiles}).
As the degree of straggling grows with the amount of material a projectile
passes through,
we expect it to be largest in our thickest film,
a detailed confirmed by simulations
(see \Cref{sec:transmitted}).
We notice that, once $\mu^{+}$ starts to penetrate into the insulating layer,
$E^{*}$ remains small (i.e., \qty{< 1}{\kilo\electronvolt})
over a wider range of $E$ for the thicker film.
We suggest that this detail,
in conjunction with the larger energy straggling,
likely means that the low-energy tail of transmitted energy distribution is
poorly encapsulated by $E^{*}$.
This is consistent with the smaller $\tilde{p}_{\ch{Mu}}$ observed for the thicker
film,
which is equivalent to a higher weighting of lower ``effective'' transmission energies.

Alternatively,
it may simply be that the differences observed in $\tilde{p}_{\ch{Mu}}(E^{*})$
are intrinsic to each film.
As mentioned above,
the muonium formation probability is known to vary somewhat depending on the
\ch{SiO2} sample
(see, e.g.,~\cite{1984-Spencer-HI-18-567,2002-Morenzoni-NIMB-192-245,2003-Prokscha-PB-326-51,2007-Prokscha-PRL-98-227401,2012-Antognini-PRL-108-143401}).
In the high-$E$ limit,
our $\tilde{p}_{\ch{Mu}}^{\infty}$s agree nicely with the expected range of variation
for the insulator,
so the differences are isolated to low $E^{*}$.
While it is difficult to be conclusive about their origin,
we point out that the \emph{average} behavior $\langle \tilde{p} \rangle (E^{*})$
is in good accord with the $E$-dependence from other
measurements in \ch{SiO2}~\cite{2002-Morenzoni-NIMB-192-245,2003-Prokscha-PB-326-51,2007-Prokscha-PRL-98-227401}.
Thus,
the overall behavior is consistent with what is known about $\mu^{+}$ in the
insulating layer.
In the future,
it may be interesting to explore this energy-dependence across a variety sample
sources.

One important aspect not considered in detail thus far is the impact
of $\mu^{+}$ \emph{diffusion} in \ch{Nb} (see \Cref{sec:results}).
While $\mu^{+}$ can viewed chemically as a \emph{very}
light proton (see, e.g.,~\cite{2024-Fleming-MSSMACMS}),
its small mass makes is much more amenable to translational motion,
leading to substantial mobility in many metals~\cite{2014-Karlsson-EPJH-39-303},
even at at cryogenic temperatures~\cite{1998-Storchak-RMP-70-929}.
\ch{Nb} is no exception,
with
the exponential decay of $P_{\mu}(t)$ implying $\mu^{+}$'s mobility at \qty{200}{\kelvin}.
In the static limit,
$P_{\mu}(t)$'s envelope is expected to be Gaussian,
reflecting a normal distribution of fields at the stopping site.
Quite generally,
when the local field is altered by a stochastic process
(e.g., thermally activated site-to-site probe ``hopping''),
the field distribution is dynamically modified.
When the rate $\nu$ of the dynamics is sufficiently high
(i.e., $\nu \gtrsim \omega_{\mu}$),
dynamic averaging causes the field distribution to ``narrow'' to a Lorentzian,
yielding an exponential envelope with a decay constant given by
(see, e.g.,~\cite{2024-Amato-IMSS}):
\begin{equation}
	\label{eq:tf-rate}
	\lambda = \frac{\sigma^{2}}{\nu} \left ( 1 + \frac{1}{1 + (\omega_{\mu} / \nu)^{2}} \right ) ,
\end{equation}
where $\sigma$ is the (static-limit) Gaussian damping rate.
Noting the small $E$-dependence to $\lambda$ in \Cref{fig:fitpar}
(i.e., from different values in the \ch{Nb} and \ch{SiO2} layers),
we take \ch{Nb}'s measured asymptotic value
$\lambda_{\ch{Nb}} = \qty{0.1039 \pm 0.0006}{\per\micro\second}$,
along with
$\sigma \approx \qty{0.50 \pm 0.02}{\per\micro\second}$
(see, e.g.,~\cite{1982-Boekema-PRB-26-2341,2013-Maisuradze-PRB-88-140509,2017-Junginger-SST-30-125013})
and
$\omega_{\mu} = \qty{8.6585 \pm 0.0013}{\per\micro\second}$,
which from \Cref{eq:tf-rate} yield
$\nu = \qty{2.61 \pm 0.24}{\per\micro\second}$~\footnote{For comparison, had we assumed our measurement was in the extreme motional-narrowing limit, \Cref{eq:tf-rate} reduces to $\lambda = 2\sigma^{2}/\nu$ (see, e.g.,~\cite{2024-Amato-IMSS}) and $\nu = \qty{4.8 \pm 0.4}{\per\micro\second}$, which differs from the more accurate estimate by \num{1.84 \pm 0.23}.}.
This is much less than the $\nu \gtrsim \qty{e9}{\per\second}$ criteria~\cite{2002-Morenzoni-NIMB-192-245}
needed to appreciably influence the $A_{0,i}$s
reported in \Cref{fig:fitpar}.
For comparison,
if we assume that $\mu^{+}$ migration takes place via high-symmetry interstitial sites
(e.g., the ``octahedral'' or ``tetrahedral'' positions)~\cite{1965-Beshers-JAP-36-290},
this corresponds to a diffusion length on the order of \qty{\sim 0.5}{\nano\meter} over $\mu^{+}$'s lifetime,
which is negligible.
Therefore,
we reiterate the assertion made in \Cref{sec:results}
that this modest mobility is insufficient to be of
detriment to the implantation measurements.

Having considered the details of our stopping profile comparisons,
we conclude this section with a brief discussion of the impact of our
results.
In order to accurately track the length scales probed by \gls{le-musr},
it is essential that the stopping simulations used to complement the
experimental work provide reliable estimates for $\mu^{+}$'s range.
This need is even more imperative when depth-resolved information
are at the heart of the physics under investigation
(e.g., the length scales associated with a superconducting state~\cite{2023-McFadden-PRA-19-044018,2024-Asaduzzaman-SST-37-025002}). 
This accuracy has been confirmed for many ``noble'' metals~\cite{2000-Gluckler-PB-289-658,2002-Morenzoni-NIMB-192-245} commonly used,
for example,
as substrates, spacers, or capping layers in condensed matter research;
however,
the available stopping cross section data~\cite{2017-Montanari-NIMB-408-50,2024-Montanari-NIMB-551-165336}
suggest this agreement is less established across the periodic table.
The present case of \ch{Nb} is exemplary of this point,
with earlier measurements having likely underestimated its superconducting
length scales (e.g., its magnetic penetration depth)~\cite{2005-Suter-PRB-72-024506,2017-Junginger-SST-30-125013,2014-Romanenko-APL-104-072601}.
In the future,
it would be interesting to perform similar experiments to those described herein
on other elemental metals where stopping data is either sparse
(e.g., \ch{Na}, \ch{Ru}, or \ch{Eu})
or
absent
(e.g., \ch{Ho}, \ch{Rh}, or \ch{Pr})~\cite{2017-Montanari-NIMB-408-50,2024-Montanari-NIMB-551-165336}.
Such an undertaking may prove challenging in rare-earth metals
due to their strong electronic moments,
which may be magnetically ordered.
In the intervening time before such experimental verifications are made,
leveraging the available stopping data
and
applying contemporary predictive techniques 
(see, e.g.,~\cite{2020-Parfitt-NIMB-478-21,2022-BivortHaiek-JAP-132-245103,2023-Akbari-NIMB-538-8,2024-Minagawa-NIMB-553-165383})
may be the best approach for ensuring accurate depth-resolution with \gls{le-musr}.

\section{
	Conclusions
	\label{sec:conclusions}
}

Using \gls{le-musr},
we studied the range of $E = \qtyrange{1.3}{23.3}{\kilo\electronvolt}$ implanted
$\mu^{+}$ in
\ch{Nb2O5}($x$~\unit{\nano\meter})/\ch{Nb}($y$~\unit{\nano\meter})/\ch{SiO2}
[$x = \qty{3.6}{\nano\meter}, \qty{3.3}{\nano\meter}$; $y = \qty{42.0}{\nano\meter}, \qty{60.1}{\nano\meter}$]
films
using the observed $\mu^{+}$ fraction stopped in a diamagnetic environment.
Using the Monte Carlo code \texttt{TRIM.SP}~\cite{1984-Biersack-APA-34-73,1991-Eckstein-SSMS-10,1994-Eckstein-REDS-1-239},
treating $\mu^{+}$ as a light $p^{+}$
with simulation inputs tuned for \gls{le-musr}~\cite{2002-Morenzoni-NIMB-192-245},
we compared the measured fractions against predictions
using our revised
and
older tabulations~\cite{1977-Anderson-SRIM-3,1993-ICRU-49}
of $p^{+}$ stopping powers in \ch{Nb}.
In conjunction with a model for \ch{Mu}'s $E$-dependent formation probability,
we find that revised stopping coefficients
yield predictions in excellent agreement with the \gls{le-musr} measurement,
while the older tabulations~\cite{1977-Anderson-SRIM-3,1993-ICRU-49}
underestimate $\mu^{+}$'s range in \ch{Nb}.
Our results indicate that earlier experiments
---
with a few exceptions~\cite{2023-McFadden-PRA-19-044018,2024-Asaduzzaman-SST-37-085006}
---
relying on
knowledge of the $\mu^{+}$ stopping profile in \ch{Nb} or \ch{Nb} compounds 
(e.g., measurements of the Meissner screening profile)
likely underestimate the lengths derived in their work.
In the future,
it would be interesting to perform similar tests on
other elements in the stopping power database~\cite{2017-Montanari-NIMB-408-50,2024-Montanari-NIMB-551-165336},
especially those where low-energy data is sparse or absent.

\begin{acknowledgments}
	T.~Junginger acknowledges financial support from \acrshort{nserc}.
	This work is based on experiments performed at the \gls{sms},
	\gls{psi}, Villigen, Switzerland (proposal number 20230035).
\end{acknowledgments}

\newpage

\appendix

\section{
	Film Thickness characterization
	\label{sec:thickness}
}

\begin{figure}
	\centering
	\includegraphics[width=1.0\columnwidth]{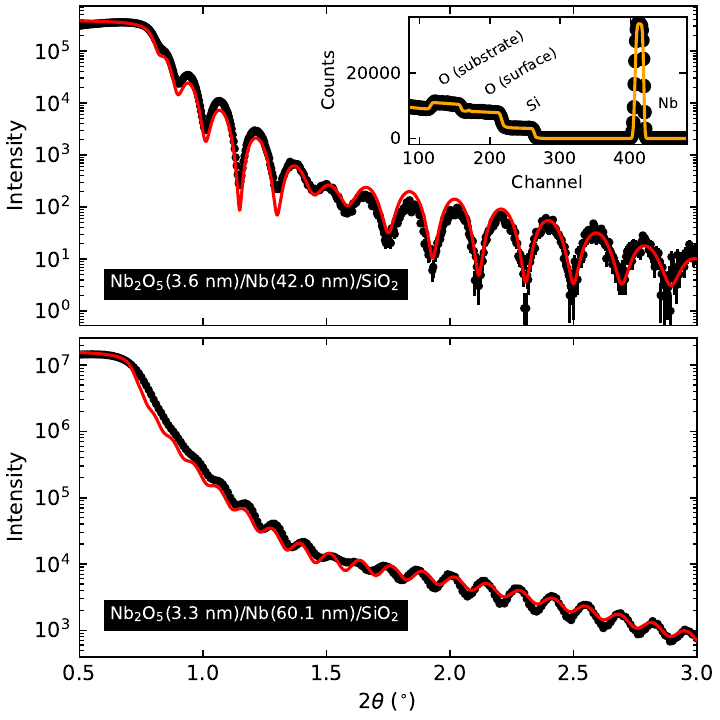}
	\caption{
		\label{fig:xrr}
		\gls{xrr} measurements of the two \ch{Nb} films.
		The solid red lines are fits to the data,
		with the corresponding layer thicknesses indicated in
		each subplot and \Cref{tab:film-thickness}.
		As an independent check,
		\gls{rbs} measurements on the thinner film were also performed
		(data and \texttt{SIMNRA}~\cite{1999-Mayer-AIPCP-475-541} simulation shown in the inset),
		yielding thickness estimates in good agreement with \gls{xrr}
		(see \Cref{tab:film-thickness}).
	}
\end{figure}

As mentioned in \Cref{sec:experiment:samples},
the thickness of each Nb film was confirmed
using \gls{xrr}.
Reflectivity spectra in both films and a fit to the data are shown in \Cref{fig:xrr},
with the corresponding layer thicknesses tabulated in \Cref{tab:film-thickness}.
As an independent check,
\gls{rbs} measurements
(\qty{2}{\mega\electronvolt} \ch{^{4}He^{+}} beam
with a scattering angle of \qty{167.5}{\degree} and a \qty{30}{\degree} tilt)
were performed on one of the films (see \Cref{fig:xrr}),
yielding thickness estimates in good agreement with \gls{xrr}
(see \Cref{tab:film-thickness}).
\Gls{erda} measurements
(ToF-E spectrometer with \qty{13}{\mega\electronvolt} \ch{^{127}I} under a total scattering angle of \qty{36}{\degree}~\cite{2005-Dobbel-NIMB-241-428})
were also performed on this film,
confirming the absence of contamination by light elements
(e.g., \ch{H}, \ch{C}, and \ch{N}).

\begin{table}
	\caption{
		\label{tab:film-thickness}
		Summary of the layer thickness measurements using \gls{xrr} and \gls{rbs}
		for the two
		\ch{Nb2O5}($x$ \unit{\nano\meter})/\ch{Nb}($y$ \unit{\nano\meter})/\ch{SiO2}(\qty{300}{\nano\meter})/\ch{Si}
		films.
	}
	\begin{ruledtabular}
	\begin{tabular}{l S S l}
		Sample & {$x$ (\unit{\nano\meter})} & {$y$ (\unit{\nano\meter})} & Method \\
		\hline
		Thin   &  3.6         & 42.0         & \gls{xrr} \\
		       &  3.6 \pm 0.4 & 45.5 \pm 2.3 & \gls{rbs} \\[1ex]
		Thick  &  3.3         & 60.1         & \gls{xrr} \\
	\end{tabular}
	\end{ruledtabular}
\end{table}

\section{
	Electronic Stopping Cross Sections
	\label{sec:cross-sections}
}

\begin{figure}
	\centering
	\includegraphics[width=1.0\columnwidth]{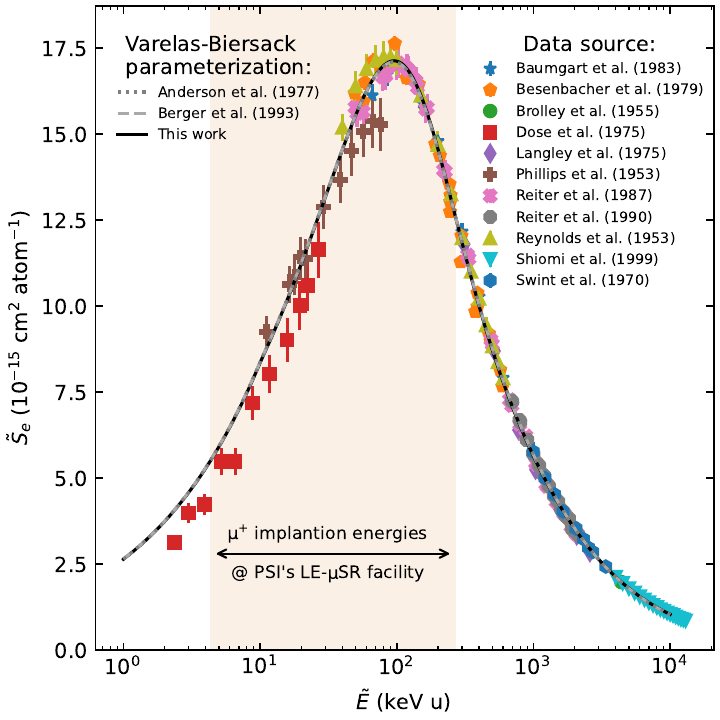}
	\caption{
		\label{fig:stopping-cross-sections-O}
		Measured electronic stopping cross sections $\tilde{S}_{e}$ for proton-like projectiles implanted in
		\ch{O2} at different scaled energies $\tilde{E}$.
		The coloured points denote results from all available studies~\cite{1953-Phillips-PR-90-532,1953-Reynolds-PR-92-742,1955-Brolley-PR-98-1112,1970-Swint-NIM-80-134,1975-Dose-ZPA-272-237,1975-Langley-PRB-12-3575,1979-Besenbacher-MFMDVS-40-3,1983-Baumgart-NIM-204-597,1987-Reiter-NIMB-27-287,1990-Reiter-NIMB-44-399,1999-Shiomi-Tsuda-NIMB-149-17}.
		Several Varelas-Biersack~\cite{1970-Varelas-NIM-79-213} parameterizations are shown~\cite{1977-Anderson-SRIM-3,1993-ICRU-49}
		along with our own fit,
		each in good agreement with the data.  
		The data were curated by the \gls{iaea} as part of their electronic stopping
		database~\cite{2017-Montanari-NIMB-408-50,2024-Montanari-NIMB-551-165336}.
	}
\end{figure}

\begin{figure}
	\centering
	\includegraphics[width=1.0\columnwidth]{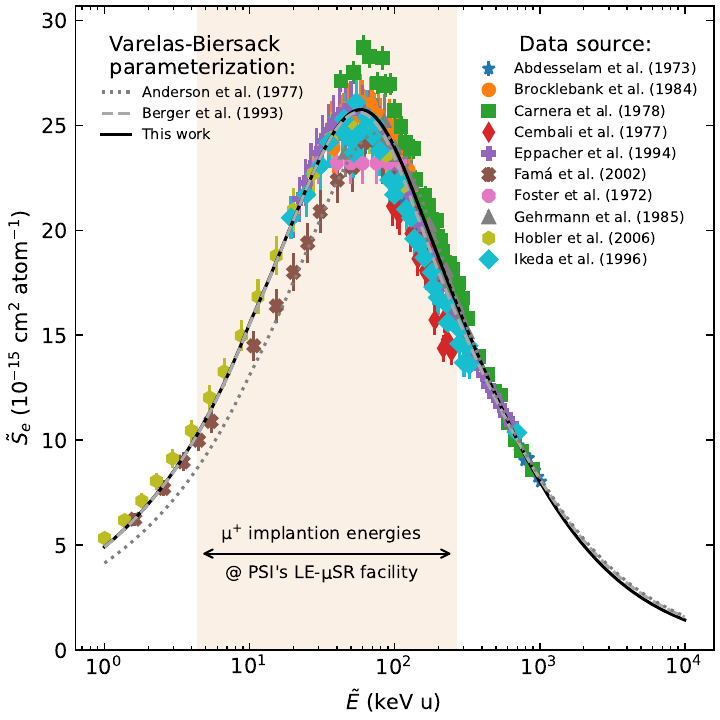}
	\caption{
		\label{fig:stopping-cross-sections-Si}
		Measured electronic stopping cross sections $\tilde{S}_{e}$ for proton-like projectiles implanted in
		\ch{Si} at different scaled energies $\tilde{E}$.
		The coloured points denote results from all available studies~\cite{1972-Foster-RE-16-139,1977-Cembali-RE-31-169,1978-Carnera-PRB-17-3492,1985-Gehrmann-PSSB-131-519,1994-Eppacher-PhD,1996-Ikeda-NIMB-115-34,2002-Fama-NIMB-193-91,2006-Hobler-NIMB-242-617,2008-Abdesselam-NIMB-266-3899,2016-Brocklebank-EPJD-70-248}
		(apart from two sets of outliers~\cite{1969-Arkhipov-SPJEPT-29-615,1976-Grahmann-NIM-132-119}).
		Several Varelas-Biersack~\cite{1970-Varelas-NIM-79-213} parameterizations are shown~\cite{1977-Anderson-SRIM-3,1993-ICRU-49}
		along with our own fit,
		each in good agreement with the data.  
		The data were curated by the \gls{iaea} as part of their electronic stopping
		database~\cite{2017-Montanari-NIMB-408-50,2024-Montanari-NIMB-551-165336}.
	}
\end{figure}

\begin{figure}
	\centering
	\includegraphics[width=1.0\columnwidth]{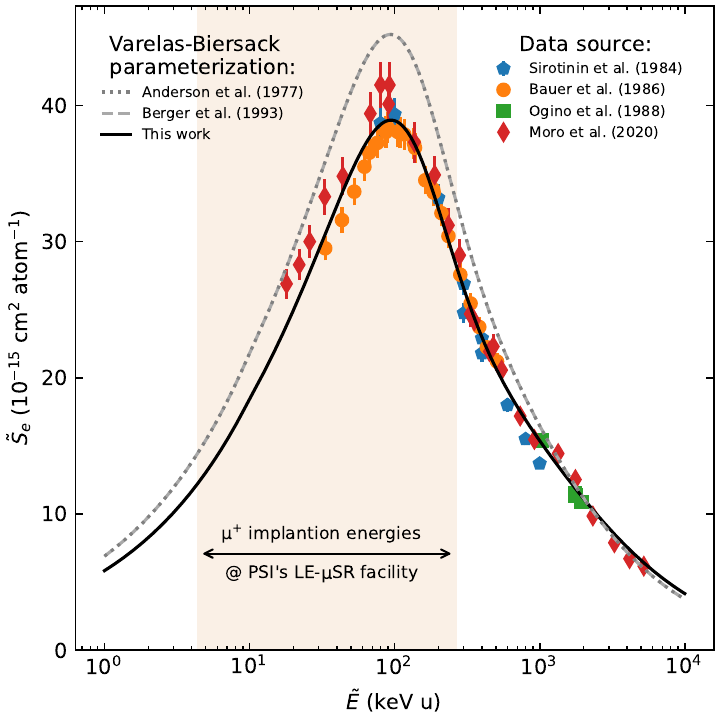}
	\caption{
		\label{fig:stopping-cross-sections-Nb}
		Measured electronic stopping cross sections $\tilde{S}_{e}$ for proton-like projectiles implanted in
		\ch{Nb} at different scaled energies $\tilde{E}$.
		The coloured points denote results from all available studies~\cite{1984-Sirotinin-NIMB-4-337, 1986-Bauer-NIMB-13-201, 1988-Ogino-NIMB-33-155, 2020-Moro-PRA-102-022808}
		(apart from one set of outliers~\cite{1973-Behr-TSF-19-247}).
		Several Varelas-Biersack~\cite{1970-Varelas-NIM-79-213} parameterizations are shown~\cite{1977-Anderson-SRIM-3,1993-ICRU-49}
		along with our own fit,
		which supersedes our previous parameterization
		(see Ref.~\cite{2023-McFadden-PRA-19-044018}).
		While our revised fit is in excellent agreement with the data,
		the older tabulations~\cite{1977-Anderson-SRIM-3,1993-ICRU-49},
		overestimate $\tilde{S}_{e}$ near the peak maximum. 
		The data were curated by the \gls{iaea} as part of their electronic stopping
		database~\cite{2017-Montanari-NIMB-408-50,2024-Montanari-NIMB-551-165336}.
	}
\end{figure}

\begin{table*}
	\centering
	\caption{
		\label{tab:coeff}
		Values for the (empirical) Varelas-Biersack~\cite{1970-Varelas-NIM-79-213} coefficients $A_{i}$
		in \Cref{eq:varelas-biersack,eq:s-low,eq:s-high}
		parameterizing the electronic stopping cross section $\tilde{S}_{e}$
		for proton-like projectiles in different elemental targets.
		Note that $\unit{\stoppingunit} \equiv \qty{e-15}{\electronvolt\centi\meter\squared\per\atom}$
		has been used when defining the units for some of the $A_{i}$s.
	}
	\begin{tabular*}{\textwidth}{l @{\extracolsep{\fill}} S S S S S l}
\hline
Target & {$A_{1}$ (\unit{\stoppingunit\kilo\electronvolt\tothe{-1/2}})} & {$A_{2}$ (\unit{\stoppingunit\kilo\electronvolt\tothe{-0.45}})} & {$A_{3}$ (\unit{\stoppingunit\kilo\electronvolt})} & {$A_{4}$ (\unit{\kilo\electronvolt})} & {$A_{5}$ (\unit{\per\kilo\electronvolt})} & Ref.\\
\botrule
\ch{O2} & 2.652 & 3 & 1920 & 2000 & 0.0223 & \cite{1977-Anderson-SRIM-3} \\
 & 2.652 & 3 & 1920 & 2000 & 0.0223 & \cite{1993-ICRU-49} \\
 & 2.645 \pm 0.014 & 2.992 \pm 0.016 & 1984 \pm 16 & 1900 \pm 60 & 0.0195 \pm 0.0007 & This work \\[1em]
\ch{Si} & 4.15 & 4.7 & 3329 & 550 & 0.01321 & \cite{1977-Anderson-SRIM-3} \\
 & 4.914 & 5.598 & 3193 & 232.7 & 0.01419 & \cite{1993-ICRU-49} \\
 & 4.92 \pm 0.04 & 5.61 \pm 0.05 & 2610 \pm 100 & 370 \pm 27 & 0.0253 \pm 0.0028 & This work \\[1em]
\ch{Nb} & 6.902 & 7.791 & 9333 & 442.7 & 0.005587 & \cite{1977-Anderson-SRIM-3} \\
 & 6.901 & 7.791 & 9333 & 442.7 & 0.005587 & \cite{1993-ICRU-49} \\
 & 5.84 \pm 0.06 & 6.58 \pm 0.07 & 12600 \pm 600 & 206 \pm 11 & 0.00270 \pm 0.00024 & This work \\
\botrule
\end{tabular*}

\end{table*}

Critical to accurate projectile range estimation from ion-implantation
simulation codes,
such as the \gls{bca} Monte Carlo application \texttt{TRIM.SP}~\cite{1984-Biersack-APA-34-73,1991-Eckstein-SSMS-10,1994-Eckstein-REDS-1-239}
used here
(see \Cref{sec:experiment:trimsp}),
is the handling of the electronic contribution to the target material's total stopping power.
For proton-like projectiles
(e.g., $\mu^{+}$),
this (inelastic) energy dissipation
can be treated empirically using the Varelas-Biersack expressions~\cite{1970-Varelas-NIM-79-213},
which parameterize an elemental target's electronic stopping cross sections $\tilde{S}_{e}$,
in terms of five coefficients $A_{i}$ ($i \in \{1,2,3,4,5\}$)
that cover the ``low-energy'' implantation region near $\tilde{S}_{e}$'s maximum.
the $E$-dependence is expressed in terms of the scaled quantity:
\begin{equation}
	\label{eq:reduced-energy}
	\tilde{E} \equiv E \left ( \frac{ \mathrm{u} }{ m } \right ) ,
\end{equation}
where $m$ is the projectile's mass
and
\unit{\amu} is the atomic mass unit~\cite{2021-Tiesinga-RMP-93-025010,1993-ICRU-49},
the latter ensuring both $E$ and $\tilde{E}$ are dimensionally equivalent
(cf.~\cite{1977-Anderson-SRIM-3}).
Explicitly,
$\tilde{S}_{e}$ is given by~\cite{1970-Varelas-NIM-79-213,1977-Anderson-SRIM-3,1993-ICRU-49}:
\begin{equation}
	\label{eq:varelas-biersack}
	\tilde{S}_{e} = \begin{cases} 
		 A_{1} \sqrt{\tilde{E}} , & \tilde{E} \leq \qty{10}{\kilo\electronvolt}, \\
		\dfrac{ s_{\mathrm{low}}(\tilde{E}) \, s_{\mathrm{high}}(\tilde{E}) }{ s_{\mathrm{low}}(\tilde{E}) + s_{\mathrm{high}}(\tilde{E}) } , & \tilde{E} > \qty{10}{\kilo\electronvolt} , \\
   \end{cases}
\end{equation}
where
\begin{equation}
	\label{eq:s-low}
	s_{\mathrm{low}}(\tilde{E}) = A_{2} \tilde{E}^{0.45} ,
\end{equation}
and 
\begin{equation}
	\label{eq:s-high}
	s_{\mathrm{high}}(\tilde{E}) = \left ( \frac{ A_{3} }{ \tilde{E} } \right ) \ln \left ( 1 + \frac{A_{4}}{\tilde{E}} + A_{5} \tilde{E} \right ) .
\end{equation}
While these expressions are not unique
(see, e.g.,~\cite{srim}),
they provide an excellent description of $\tilde{S}_{e}$'s $\tilde{E}$-dependence
in most elements.
For compound targets,
Bragg's additivity rule~\cite{1905-Bragg-LEDPMJS-10-318} is used,
which remains a reasonable approximation at low $\tilde{E}$~\cite{2024-Montanari-NIMB-551-165336}.

While the qualitative form of \Cref{eq:varelas-biersack,eq:s-low,eq:s-high}
is known to be correct,
their accuracy depends on the $A_{i}$s used,
with tabulations for target elements up to uranium available~\cite{1977-Anderson-SRIM-3,1993-ICRU-49}.
Recently,
the validity of these values for \ch{Nb} was considered~\cite{2023-McFadden-PRA-19-044018},
with data from the \gls{iaea}['s] electronic stopping power database
(version 2021-12)~\cite{2017-Montanari-NIMB-408-50,2024-Montanari-NIMB-551-165336}
showing that the tabulations~\cite{1977-Anderson-SRIM-3,1993-ICRU-49}
overestimate $\tilde{S}_{e}$ around its maximum.
Here,
we expand on this assessment using the database's
latest version (2024-03)~\cite{2017-Montanari-NIMB-408-50,2024-Montanari-NIMB-551-165336}
and
include additional elemental targets.

\Cref{fig:stopping-cross-sections-O,fig:stopping-cross-sections-Si,fig:stopping-cross-sections-Nb}
show measured $\tilde{S}_{e}$s for proton-like projectiles implanted in
\ch{O2}~\cite{1953-Phillips-PR-90-532,1953-Reynolds-PR-92-742,1955-Brolley-PR-98-1112,1970-Swint-NIM-80-134,1975-Dose-ZPA-272-237,1975-Langley-PRB-12-3575,1979-Besenbacher-MFMDVS-40-3,1983-Baumgart-NIM-204-597,1987-Reiter-NIMB-27-287,1990-Reiter-NIMB-44-399,1999-Shiomi-Tsuda-NIMB-149-17},
\ch{Si}~\cite{1972-Foster-RE-16-139,1977-Cembali-RE-31-169,1978-Carnera-PRB-17-3492,1985-Gehrmann-PSSB-131-519,1994-Eppacher-PhD,1996-Ikeda-NIMB-115-34,2002-Fama-NIMB-193-91,2006-Hobler-NIMB-242-617,2008-Abdesselam-NIMB-266-3899,2016-Brocklebank-EPJD-70-248},
and
\ch{Nb}~\cite{1984-Sirotinin-NIMB-4-337, 1986-Bauer-NIMB-13-201, 1988-Ogino-NIMB-33-155, 2020-Moro-PRA-102-022808}
targets.
Note that measurements which are clear outliers
(Refs.~\cite{1969-Arkhipov-SPJEPT-29-615,1976-Grahmann-NIM-132-119} for \ch{Si}
and
Ref.~\cite{1973-Behr-TSF-19-247} for \ch{Nb})
have been omitted.
For each target's data,
we fit the $\tilde{E}$-dependence using
the Varelas-Biersack expressions~\cite{1970-Varelas-NIM-79-213}, 
with the result shown as a solid black line.
In doing this,
we followed the approach in Ref.~\cite{2023-McFadden-PRA-19-044018},
where it was pointed out that the smooth continuity of
\Cref{eq:varelas-biersack}
implies that:
\begin{equation*}
	A_{1} \equiv \left ( \frac{1}{\sqrt{ \qty{10}{\kilo\electronvolt} }} \right ) \frac{ s_{\mathrm{low}}( \qty{10}{\kilo\electronvolt} ) \, s_{\mathrm{high}}( \qty{10}{\kilo\electronvolt} ) }{ s_{\mathrm{low}}( \qty{10}{\kilo\electronvolt} ) + s_{\mathrm{high}}( \qty{10}{\kilo\electronvolt} ) } ,
\end{equation*}
reducing the number fitted $A_{i}$s from five to four.
This choice is most consequential for fitting the $\tilde{S}_{e}$ data in \ch{Nb}
(see \Cref{fig:stopping-cross-sections-Nb}),
where measurements in the region $\tilde{E} \lesssim \qty{10}{\kilo\electronvolt}$
are absent.
For \ch{O2} (see \Cref{fig:stopping-cross-sections-O})
and
\ch{Si} (see \Cref{fig:stopping-cross-sections-Si}),
this constraint is not detrimental to the fit,
as evidenced by the good agreement with the data in this $\tilde{E}$-region.
For comparison,
$\tilde{S}_{e}(\tilde{E})$ curves using literature $A_{i}$s~\cite{1977-Anderson-SRIM-3,1993-ICRU-49}
are also shown in \Cref{fig:stopping-cross-sections-O,fig:stopping-cross-sections-Si,fig:stopping-cross-sections-Nb}.
In each case,
our fit gives the best agreement with the $\tilde{S}_{e}$ data.
Explicit values for our newly determined $A_{i}$s,
along with values for the older tabulations~\cite{1977-Anderson-SRIM-3,1993-ICRU-49},
are given in \Cref{tab:coeff}.
Note that the values reported in \Cref{tab:coeff} for \ch{Nb} supersede those in Ref.~\cite{2023-McFadden-PRA-19-044018},
which were derived from an earlier version (2021-12) of the \gls{iaea}['s] database~\cite{2017-Montanari-NIMB-408-50,2024-Montanari-NIMB-551-165336}
that did not explicitly include uncertainty estimates.

\section{
	$\mu^{+}$ Transmission through \ch{Nb2O5}/\ch{Nb}
	\label{sec:transmitted}
}

\begin{figure}
	\centering
	\includegraphics[width=1.0\columnwidth]{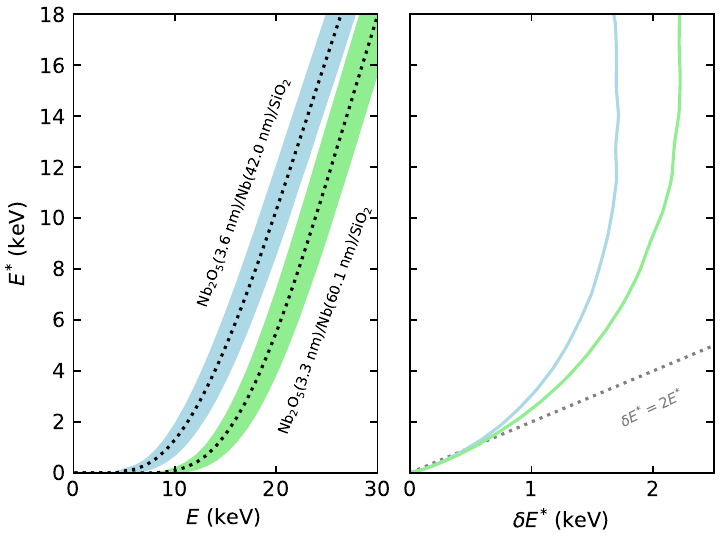}
	\caption{
		\label{fig:etrans}
		Mean energy $E^{*}$ of $\mu^{+}$ transmitted through the
		\ch{Nb2O5}/\ch{Nb} layers as a function of implantation
		energy $E$,
		obtained from \texttt{TRIM.SP}~\cite{1984-Biersack-APA-34-73,1991-Eckstein-SSMS-10,1994-Eckstein-REDS-1-239}
		simulations without the \ch{SiO2} substrate.
		The shaded colored regions denote the span
		of $E^{*} \pm \delta E^{*}$ (its standard deviation).
		The relationship between 
		$E^{*}$ and $\delta E^{*}$ is also shown.
		In these simulations,
		we have used the Varelas-Biersack~\cite{1970-Varelas-NIM-79-213}
		coefficients derived in this work as inputs
		(see \Cref{tab:coeff}).
	}
\end{figure}

In the model introduced in \Cref{sec:results} to account for \ch{Mu}'s
energy-dependent formation probability [\Cref{eq:p_Mu}],
one of its dependencies is the mean energy $E^{*}$ of $\mu^{+}$
\emph{transmitted} through the \ch{Nb2O5}/\ch{Nb} layers.
This quantity is not directly available from implantation simulations
shown in \Cref{fig:implantation-profiles},
but may be obtained from separate
\texttt{TRIM.SP}~\cite{1984-Biersack-APA-34-73,1991-Eckstein-SSMS-10,1994-Eckstein-REDS-1-239}
calculations where the \ch{SiO2} substrate is omitted.
The $E$-dependence of $E^{*}$ for our two films is shown in \Cref{fig:etrans},
where we have used the Varelas-Biersack~\cite{1970-Varelas-NIM-79-213}
coefficients derived in this work as inputs (see \Cref{tab:coeff}).
While an $E^{*} \approx \qty{0}{\kilo\electronvolt}$ indicates
insufficient passage into the insulating layer,
$E^{*}$ increases monotonically
once $E$ is sufficiently high.
Importantly,
we point out that encapsulating $\mu^{+}$ transmission using $E^{*}$ is an approximation;
the energy of muons eluting out of the \ch{Nb2O5}/\ch{Nb} layers really
follow a \emph{distribution}
(i.e., due to range straggling).
This is emphasized by he shaded colored regions
in \Cref{fig:etrans},
which denote the span of $E^{*} \pm$ its standard deviation $\delta E^{*}$.
Notice that,
particularly for the thicker film,
the $E^{*} \pm \delta E^{*}$ ``band'' remains close to zero
for a wider $E$-region than in the thinner film.
This detail,
in conjunction with the larger $\delta E^{*}$ for the thicker film,
likely contributes to the different $\tilde{p}_{\ch{Mu}}(E^{*})$s
observed for the two films 
at low $E^{*}$ values
(see \Cref{fig:mu-formation}).

\bibliography{references.bib,unpublished.bib}

\end{document}